\documentclass[%
 reprint,
 amsmath,amssymb,
 aps,longbibliography,
pra,
superscriptaddress
]{revtex4-2}

\usepackage[utf8]{inputenc}
\usepackage{graphicx}
\usepackage{bm}
\usepackage{bbold}
\usepackage{braket}
\usepackage{tikz,pgfplots}
\usepackage{xcolor}
\usepackage{tabularx}
\usepackage{booktabs}
\usepackage[outdir=./]{epstopdf}
\pgfplotsset{compat=1.10}
\usepackage{siunitx}
\usepackage[version=4]{mhchem}

\DeclareMathOperator*{\SumInt}{%
\mathchoice%
  {\ooalign{$\displaystyle\sum$\cr\hidewidth$\displaystyle\int$\hidewidth\cr}}
  {\ooalign{\raisebox{.14\height}{\scalebox{.7}{$\textstyle\sum$}}\cr\hidewidth$\textstyle\int$\hidewidth\cr}}
  {\ooalign{\raisebox{.2\height}{\scalebox{.6}{$\scriptstyle\sum$}}\cr$\scriptstyle\int$\cr}}
  {\ooalign{\raisebox{.2\height}{\scalebox{.6}{$\scriptstyle\sum$}}\cr$\scriptstyle\int$\cr}}
}

\usetikzlibrary{patterns,shadows,fadings,positioning,trees,calc}
\usetikzlibrary{shapes}

\DeclareMathOperator\erf{erf}

\tikzfading[name=fade cloud,
         inner color=transparent!99,
         outer color=transparent!40]
\tikzfading[name=fade inside,
         inner color=transparent!80,
         outer color=transparent!30]
\tikzfading[name=fade out,
         inner color=transparent!0,
         outer color=transparent!90]
\tikzfading[name=fade inner,
         inner color=transparent!100,
         outer color=transparent!0]

\definecolor{diplom1}{RGB}{101 156 239}
\definecolor{diplom2}{RGB}{000 000 128}
\definecolor{diplom3}{RGB}{153,0,0} 
\definecolor{diplom4}{RGB}{232,215,23}
\definecolor{diplom5}{RGB}{51,37,60}

\definecolor{unirot}{RGB}{153,0,0}
\definecolor{unirot_hell}{RGB}{255,228,225}
\definecolor{lightblue}{RGB}{242.2,249.88,255}

\AtBeginDocument{
\heavyrulewidth=.08em
\lightrulewidth=.05em
\cmidrulewidth=.03em
\belowrulesep=.65ex
\belowbottomsep=0pt
\aboverulesep=.4ex
\abovetopsep=0pt
\cmidrulesep=\doublerulesep
\cmidrulekern=.5em
\defaultaddspace=.5em
}

\usepackage{hyperref}

\begin{document}


\title{Unravelling time-resolved Interparticle Coulombic Decay: From spectral formation to decay lifetimes}

\author{Alexander V. Riegel}%
\affiliation{%
Center for Light--Matter Interaction, Sensors \& Analytics (LISA$^+$),
University of Tübingen\\
Auf der Morgenstelle 15, 72076 Tübingen, Germany
}%
\affiliation{%
Institute of Physical and Theoretical Chemistry,
University of Tübingen\\
Auf der Morgenstelle 18, 72076 Tübingen, Germany
}%
\author{Max K. Humm}%
\affiliation{%
Institute of Physical and Theoretical Chemistry,
University of Tübingen\\
Auf der Morgenstelle 18, 72076 Tübingen, Germany
}%
\author{Jan-Thore Kahle}%
\affiliation{%
Institute of Physical and Theoretical Chemistry,
University of Tübingen\\
Auf der Morgenstelle 18, 72076 Tübingen, Germany
}%
\author{Elke Fasshauer}
 \email{elke.fasshauer@pm.me}
\affiliation{%
Center for Light--Matter Interaction, Sensors \& Analytics (LISA$^+$),
University of Tübingen\\
Auf der Morgenstelle 15, 72076 Tübingen, Germany
}%
\affiliation{%
Institute of Physical and Theoretical Chemistry,
University of Tübingen\\
Auf der Morgenstelle 18, 72076 Tübingen, Germany
}%

\date{\today}

\begin{abstract}
Electronic decay processes provide a fascinating window into correlated
electronic rearrangements occurring on ultrafast timescales.
Following these dynamics in real time has become increasingly
accessible experimentally, but extracting the underlying electronic
dynamics from measured spectra requires understanding how nuclear
motion shapes the observable signal.
Here, we extend an analytical description of time-resolved electronic
decay spectra to include dissociative nuclear dynamics and apply it to
Interparticle Coulombic Decay (ICD) in the neon dimer.
The resulting spectra reproduce the experimentally observed spectral
shape and reveal an unexpectedly important role of interference
between pathways involving different vibronic resonance states.
We further establish a clear connection between the temporal build-up
of the spectral structure and the nuclear wavepacket dynamics in the
decaying electronic state.
Most strikingly, our analytical expressions reveal that the
time-dependent integrated ICD signal contains contributions
proportional to both $\exp(-t/\tau)$ and $\exp[-t/(2\tau)]$.
Nevertheless, a conventional mono-exponential fit can describe the
temporal signal remarkably well while yielding a decay lifetime that
differs substantially from the underlying value.
Applying the theoretically derived fitting model to experimental data
for the neon dimer yields an ICD lifetime $\tau$ of \qty{73}{\fs}, rather than
the previously extracted
\qty[separate-uncertainty = true]{150(50)}{\fs}, placing the
experimental value within the range of previous theoretical
predictions.
Since the underlying temporal structure is common to electronic decay
processes, our findings have implications for extracting lifetimes from
time-resolved decay spectra well beyond ICD.
\end{abstract}

\maketitle

\section{Introduction}
\label{sec:Introduction}

Excited electronic states and their decay are central to many processes
in physics and chemistry
\cite{Gonzalez2020,Balzani2024,Chulkov2006,Bhuyan2023,Knepp2025,Gallmann2017,Matsika2018,Garcia2023,Muller2025,Janos2026}.
While relaxation may occur radiatively through the emission of a
photon, excess energy can also be released through electronic decay
processes, in which one or more electrons are emitted.
Prominent examples are Auger--Meitner decay
\cite{Meitner22,Auger23} and Interparticle Coulombic Decay (ICD)
\cite{Cederbaum97,Marburger03,Jahnke04}, which can proceed on atto- to femtosecond timescales \cite{Jahnke20,Haynes2021,Hofierka2026}.
The ultrafast redistribution of energy and the correlated electronic
rearrangements underlying these processes make their dynamics
particularly fascinating.

ICD is a non-local electronic decay process that may, for example, be
initiated by the photoionization of an entity $\mathrm{A}$, such as an
atom, molecule or quantum dot, from an inner-valence or core level.
The resulting vacancy is filled by an electron of the same unit, while
the released energy is simultaneously transferred to a neighbouring
entity $\mathrm{B}$, where it leads to the emission of a secondary
electron, the ICD electron $\mathrm{e}_\text{ICD}^-$:
\begin{equation*}
\mathrm{AB} \xrightarrow{h\nu} \mathrm{A}^{**+}\mathrm{B} + \mathrm{e}_\text{ph}^- \rightarrow \mathrm{A}^{+} + \mathrm{B}^+ + \mathrm{e}_\text{ph}^- + \mathrm{e}_\text{ICD}^- \, .
\end{equation*}

In the final state of the process above, both entities carry a net
positive charge and therefore repel each other, generally leading to
dissociation through a Coulomb explosion \cite{Jahnke20,Hoener2008}.
Since its prediction, ICD has been found to occur ubiquitously in a
wide variety of systems and environments \cite{Jahnke20}, including
water clusters \cite{Mueller06,Hergenhahn11,Stoychev11,Richter18,Ren2023,Zhang2025}, quantum dots
\cite{Bande11,Dolbundalchok16,Weber2017,Agueny2020,Marando2025,Marando2026} and biomolecules \cite{Harbach13,Hans2021,Skitnevskaya2023,Gao2025}.

The continuing development of ultrafast and attosecond techniques
provides increasing opportunities to follow electronic rearrangements
directly in time \cite{AgostiniNobel,KrauszNobel,LHuillierNobel}.
Time- and energy-resolved electron spectra are particularly valuable in
this context because they can reveal how the electronic decay evolves
while nuclear degrees of freedom shape the measured signal.
An early time-resolved experiment on ICD in the neon dimer employed a
60-fs ionizing pulse and measured the build-up of the decay signal
\cite{Schnorr15}.
With increasing temporal resolution, a theoretical description is
needed that connects such spectra to the underlying electronic decay
while accounting for the accompanying nuclear dynamics.
Our analytical framework developed in
Refs.~\cite{Fasshauer20_1,Riegel25} provides access to the
time-dependent decay spectrum directly at any given time.
While nuclear dynamics in ICD, including dissociation, have been
treated previously \cite{Santra00_1,Scheit03,Sisourat10_2,Grull2020,Cabrera2020,Fedyk2023}, our time-resolved analytical description has so
far been restricted to bound nuclear states.
Extending it to a dissociative nuclear continuum is therefore required
for a proper description of ICD in systems such as the neon dimer.


A key quantity characterizing electronic decay processes is the decay
lifetime $\tau$.
In time-resolved experiments, it is commonly extracted by fitting the
time-dependent decay signal to a mono-exponential function of the form
$\exp(-t/\tau)$ \cite{Schnorr13,Schnorr15,Fukuzawa2019,Jahnke20}.
However, the temporal dependence of the measured observable does not
necessarily follow the population decay of the electronic resonance
state directly.
In our previous purely electronic description of resonant ICD (RICD), we
found that the analytical time dependence of the signal contains terms
that are inconsistent with identifying the time constant obtained from
such a mono-exponential fit directly with the decay lifetime
\cite{Fasshauer20_1}.
A discrepancy between the theoretically predicted lifetime and the one
extracted from a time-resolved electron spectrum has
also been observed for autoionization \cite{Busto18}, indicating that
this issue extends beyond ICD.
Establishing how the decay lifetime is encoded in a time-resolved
electronic-decay spectrum is therefore essential for extracting and
interpreting lifetimes from such measurements.


In this work, we extend our analytical time-resolved description to
include dissociative nuclear dynamics and derive the corresponding
expressions for the ICD spectrum in Sec.~\ref{sec:theory}.
The system and laser parameters for our application to the neon dimer,
together with the numerical treatment and discretization of the
dissociative continuum, are presented in
Sec.~\ref{sec:CompDetails}.
In Sec.~\ref{sec:results}, we investigate the physical mechanisms that
determine the spectral shape and its temporal evolution and establish
how the decay lifetime is encoded in the time-dependent ICD signal.
We test the resulting lifetime-extraction procedure against simulated
spectra with a known input lifetime and apply it to the experimental
data of Ref.~\cite{Schnorr15}.
We conclude our findings in Sec.~\ref{sec:conclusion}.

All equations are given in atomic units unless
specified otherwise.
\section{Theory}
\label{sec:theory}

\subsection{Analytical derivation of the ICD spectrum}
\label{subsec:derivation}

\begin{figure}[ht]
 \centering
 \includegraphics[width=0.74\columnwidth]{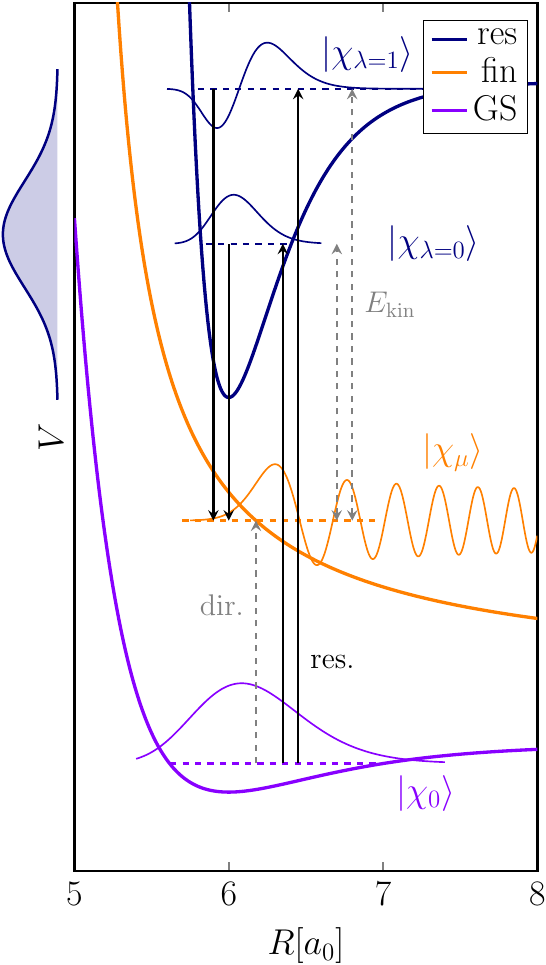}
 \caption{(Colour online) Schematic illustration \cite{figshare_pics_ICD_2026}
 of energy levels and processes in a diatomic system.
 Lower curve: electronic ground state hosting one bound vibrational state $\ket{\chi_0}$;
 upper curve: electronic resonance state hosting $N_\lambda$ bound vibrational states $\ket{\chi_{\lambda}}$;
 middle curve: dissociative electronic final state hosting a continuum of nuclear states $\ket{\chi_{\mu}}$.
 Possible pathways into a given final state:
 direct pathway (dashed arrow);
 resonance pathways (solid arrows),
 both leading to emission of electron with kinetic energy $E_\text{kin}$ (dashed double-headed arrows).
}
 \label{fig:energy_overview}
\end{figure}


The electronic decay processes we intend to describe in a fully time-dependent manner
are schematically illustrated in Fig.~\ref{fig:energy_overview}.
Starting from the ground state $\ket{G}$ of the system, a laser pulse centred around the photon energy $\Omega$ ionizes the system
and thereby creates a coherent superposition of vibrational states $\ket{\chi_{\lambda}}$ of the electronic resonance state $\ket{r}$
as well as a photoelectron with kinetic energy $E_p$.
If ICD is an allowed process, then the resonance states will decay to final states with a lifetime $\tau_{\lambda}$
by transferring parts of the wavepacket into nuclear states $\ket{\chi_{\mu}}$ of the electronic final state $\ket{E}$
under emission of a secondary ICD electron with kinetic energy $E_{\text{kin}}$.
This pathway into a final state is called "resonance pathway" and is symbolized in Fig.~\ref{fig:energy_overview} by solid arrows.
In principle, competing pathways into the same final state are possible:
the double photoionization from the ground state into the final state constituting the "direct pathway",
indicated in Fig.~\ref{fig:energy_overview} by a dashed arrow,
and an "indirect pathway" in which an intermediate state from the manifold of final states is included into the resonance pathway.

We assume a diatomic molecule with a bound electronic ground state with one vibrational state $\ket{\chi_0}$,
an electronic resonance state with $N_{\lambda}$ bound vibrational states $\ket{\chi_{\lambda}}$
and an electronic final state with a repulsive Coulomb nuclear potential
possessing a continuum of nuclear states $\ket{\chi_{\mu}}$.
Based on this model, we derive the time-resolved ICD electron spectrum as a function of the kinetic energies of the emitted photoelectron and ICD electron as well as time.


The theoretical description is largely analogous to Ref.~\cite{Riegel25}.
Our starting point is the time-dependent Schrödinger equation
\begin{equation}	\label{eq:TDSE}
	i \frac{\partial}{\partial t} \ket{\Psi(t)} = H(t) \ket{\Psi(t)}	\, .
\end{equation}
We partition the Hamiltonian
\begin{equation}	\label{eq:Hamiltonian}
	H(t) = H_\text{F} + H_{X}(t) = H_{0} + V + H_{X}(t)
\end{equation}
into two parts (see also Eqs.~(22) and (5) in Ref.~\cite{Riegel25}): 
The first one is the Fano-type Hamiltonian $H_\text{F} = H_{0} + V$ of the system unperturbed
by the laser field,
which itself can be partitioned into the Hamiltonian $H_{0}$ of the non-decaying system
and the resonance--continuum coupling operator $V$ known as configuration-interaction operator \cite{Fano61,Riegel25}.
The second one is the Hamiltonian of the light--matter interaction using length gauge and the dipole approximation, $H_{X}(t) = -\mu \mathcal{E}_X(t)$.
Here $\mu$ denotes the dipole operator and $\mathcal{E}_X$ denotes the field strength of the ionizing laser pulse.
In principle, all three dimensions of the vector-like dipole operator must be included and the transition dipole moments are usually not isotropic,
leading to the possibility of favouring or selecting certain transitions based on the orientation of the molecule and the polarization of the laser pulse \cite{Gokhberg10_1,Fasshauer13,Haller2D}.
We imagine the sample to contain molecules with random orientation and therefore expect a rotationally averaged spectrum.
Therefore and in the interest of simplicity, we focus only on the dimension along the axis containing the two nuclei. 

In order to model the ICD process, we adapt the model of Ref.~\cite{Riegel25} to include the following three properties:

\begin{enumerate}
\item In the final state, the two positively charged ions experience Coulomb repulsion from each other.
This leads to the corresponding potential-energy curve (PEC) being dissociative,
for which the nuclear final states form a continuum $\{\ket{\chi_{\mu}}\}$.
In our equations, we therefore replace all occurring summations $\sum_{\mu}$ over nuclear final states
with integrations $\int \mathrm{d}E_{\mu}$, where the nuclear final states are characterized by their energy $E_{\mu}$.

\item The difference between the processes leading to resonant ICD and to ICD
is that the moiety interacting with the laser pulse is excited in the former case but ionized in the latter case.
The ICD, therefore, produces a photoelectron with kinetic energy $E_p$
in addition to the ICD electron with kinetic energy $E_{\text{kin}}$.
As a consequence, the target quantity of our calculations is now
the probability $P$ of a two-centre double ionization as a function of the kinetic energies of both electrons and time $t$,
\begin{equation} \label{eq:td_probability}
	P(E_\text{kin}, E_p, t) = \SumInt \limits_k  \big|\braket{E_k|\Psi(t) }\big|^2	\, ,
\end{equation}
while in the analogous Eq.~(20) in Ref.~\cite{Riegel25},
the functional dependence was $P(E_\text{kin}, t)$.
In this notation, $\ket{E_k}$ is a total final continuum state with energy $E_k$
in which both kinetic energies of the electrons and the energy of the doubly ionized state $k$ are aggregated.
We will work within a product ansatz in which a total state is decomposed
into the state of the photoelectron and the state of the remaining system.
For example, the electronic state $\ket{p}$ of the photoelectron (characterized by its kinetic energy $E_p$)
and the vibronic state $\ket{E_{\text{vibron}}}$ comprising both the ICD electron and the doubly ionized residual system
will be taken together as the common product state $\ket{E_k} = \ket{p,E_{\text{vibron}}} = \ket{p} \ket{E_{\text{vibron}}}$.
Likewise, in the latter vibronic state $\ket{E_{\text{vibron}}}$ we assume the state of the ICD electron to be separable from the state of the residual system.
In doing so, we ignore the interactions of the emitted electrons in the continuum.
These may become important in certain cases, see Ref.~\cite{Ruberti24} and references therein,
and will be the subject of future work.

\item ICD and related processes are only possible because of the interaction between the different moieties.
Therefore, the effectiveness of the decay mechanism depends on their distance.
For example, the decay rate $\Gamma(R)$ as function of the internuclear distance $R$
behaves as $R^{-6}$ in the asymptotic regime for the dipole-allowed ICD \cite{Averbukh04,Gokhberg10_1,Fasshauer13},
the Electron-Transfer-Mediated Decay (ETMD) \cite{Fasshauer13}
and the proposed Interparticle Coulombic Electron Capture (ICEC) \cite{Senk24,Jahr25},
while at smaller distances, wavefunction overlap becomes more pronounced and additional terms such as exponential ones arise.
In Ref.~\cite{Riegel25} we assumed for simplicity that the decay rate $\Gamma = 2\pi |V|^2$ and the coupling operator $V$ were purely electronic.
Here we acknowledge that the decay rate, and hence the coupling strength, indeed depend on the internuclear distance.
As a consequence, matrix elements of $V$ like $\bra{\chi_{\mu}}\braket{E | V | r}\ket{\chi_{\lambda}}$ can no longer be factorized
into $\braket{E|V|r} \braket{\chi_{\mu} | \chi_{\lambda}} \equiv V_{Er} \braket{\chi_{\mu} | \chi_{\lambda}}$ (see Eq.~(A4) in Ref.~\cite{Riegel25}).
Let us assume that $V(R)$ can be written as
\begin{equation}	\label{eq:V_of_R}
	V(R) = \alpha_V \, v(R)
\end{equation}
where $v(R)$ carries only the dependence on $R$ (and is therefore diagonal in the electronic subspace),
while $\alpha_V$ is independent of $R$ (and hence diagonal in the nuclear subspace, but not necessarily in the electronic one).
Then the aforementioned matrix element can be partitioned as
\begin{equation} \label{eq:V_matrix_element}
	\bra{\chi_{\mu}}\braket{E | V | r}\ket{\chi_{\lambda}} =
	\braket{E|\alpha_V|r} \braket{\chi_{\mu} | v(R) | \chi_{\lambda}} \,.
\end{equation}
The former overlap integral of nuclear wavefunctions, $\braket{\chi_{\mu} | \chi_{\lambda}}$, has now acquired a weighting function $v(R)$ within the integral.
\end{enumerate}

Keeping these changes in mind, the course of action is similar to Ref.~\cite{Riegel25}:
In order to evaluate Eq.~\eqref{eq:td_probability} (which is analogous to Eq.~(20) in Ref.~\cite{Riegel25}), we need to find the transition amplitude
for a certain final state $\ket{p} \ket{E} \ket{\chi_{\mu}}$,
which in first-order time-dependent perturbation theory is given by (compare to Eq.~(31) in Ref.~\cite{Riegel25})
\begin{widetext}
	\begin{align}	\label{eq:trs_amplitude}
		\begin{split}
			\bra{\chi_\mu} \bra{E} \braket{p | \Psi(t)} = \,
			& -i \int\limits_{t_0}^{t} \mathrm{d}t' \,
			\bra{\chi_\mu} \bra{E} \braket{p |\, U_0(t,t') H_X(t') \,| G}\\
			& -\int\limits_{t_0}^{t} \mathrm{d}t'
			\int\limits_{t'}^t \mathrm{d}t'' \,
			\bra{\chi_\mu} \bra{E} \braket{p |\, U_0(t,t'') \, V \, U_\text{F} (t'',t') \, H_X(t') \,| G}	\, .
		\end{split}
	\end{align}
\end{widetext}
As in Ref.~\cite{Riegel25}, we have chosen the energy of the total ground state $\ket{G}$ to be $E_G = 0$.
The time-evolution operators $U_0$ and $U_{\text{F}}$ pertain to the Hamiltonians of the non-decaying system, $H_0$,
and the non-perturbed system, $H_\text{F}$, as given in Eq.~\eqref{eq:Hamiltonian}, respectively.

From a derivation analogous to the one in Ref.~\cite{Riegel25} results the following working equation (compare to Eq.~(42) in Ref.~\cite{Riegel25}):
\begin{widetext}
\begin{align}
\label{eq:ampl_working_eq}
\begin{split}
 \bra{\chi_\mu} \bra{E} \braket{p | \Psi(t)} =
                & +i \braket{c|\mu|g}
                     \int\limits_{t_0}^{t} \mathrm{d}t' \,
                     \exp \biggl[ -i (t-t') (E_\text{kin} + E_\text{fin} + E_{\mu} + E_p) \biggr] \,
                    \braket{\chi_{\mu}| \chi_0} \mathcal{E}_X(t')\\
                & + \frac{\braket{r|\mu|g}}{N_\lambda}  \sum\limits_{\lambda} 
                   \int\limits_{t_0}^{t} \mathrm{d}t'
                   \int\limits_{t'}^t \mathrm{d}t'' \,
                     \exp \biggl[ -i (t-t'') (E_\text{kin} + E_\text{fin} + E_{\mu} + E_p) \biggr]    \\
                & \qquad  \times \braket{c|\alpha_V|r} \braket{\chi_\mu | v(R) | \chi_\lambda}
                  \braket{\chi_\lambda | \chi_0}
                     \exp \biggl[ -i (t''-t') (E_r + E_\lambda + E_p - i\pi W_\lambda) \biggr] 
                  \, \mathcal{E}_X(t') \\
                & - \frac{i\pi \braket{c|\mu|g}}{N_\lambda}
                   \sum\limits_{\lambda} \int \mathrm{d}E_{\mu''}
                   \int\limits_{t_0}^{t} \mathrm{d}t'
                   \int\limits_{t'}^t \mathrm{d}t'' \,
                     \exp \biggl[ -i (t-t'') (E_\text{kin} + E_\text{fin} + E_{\mu} + E_p) \biggr]    \\
                & \qquad  \times \left|\braket{c|\alpha_V|r}\right|^2 \braket{\chi_\mu | v(R) | \chi_\lambda}
                  \braket{\chi_\lambda | v(R) | \chi_{\mu''}} \braket{\chi_{\mu''} | \chi_0}
                     \exp \biggl[ -i (t''-t') (E_r + E_\lambda + E_p - i\pi W_\lambda) \biggr] 
                  \, \mathcal{E}_X(t')  \, .\\
\end{split}
\end{align}
\end{widetext}
Here, $E_r$ and $E_\text{fin}$ denote the minima of the electronic PECs for the resonance and the final state, respectively.
$E_{\lambda}$ and $E_{\mu}$ denote the corresponding energies of the nuclear eigenstates of the resonance and the final state with respect to the according minimum.
$\braket{r|\mu|g}$ and $\braket{c|\mu|g}$ are the transition dipole moment matrix elements
for the transitions from the electronic ground state to the resonance state and to the final state, respectively.
For the latter case, we again have assumed that the matrix element is independent of the energy of the continuum final state, $\braket{E_k|\mu|g}= \braket{c|\mu|g}$,
and in both cases independent of $E_p$.
Likewise, we take the electronic coupling matrix element to be energy independent, $\braket{E_k|\alpha_V|r} = \braket{c|\alpha_V|r}$.
Finally $W_{\lambda}$ is the effective decay width from the resonance state of $\ket{\chi_{\lambda}}$;
compared to its definition in Eq.~(12) in Ref.~\cite{Riegel25} it now involves the modified coupling matrix elements, though,
$W_{\lambda} = \left|\braket{c|\alpha_V|r}\right|^2 \int \mathrm{d}E_{\mu'} \left|\braket{\chi_{\lambda} | v(R) | \chi_{\mu'}}\right|^2$,
where we have again assumed independence of the energy of the continuum final state.

Comparing Eq.~\eqref{eq:ampl_working_eq} with the working equation for the RICD with bound final states (Eq.~(42) in Ref.~\cite{Riegel25}),
we find the structure of the equations to be largely identical.
The interpretation of the three main terms as direct, resonance and indirect pathways from the ground state to the chosen final state, respectively, is still plausible.
Apart from the integral over $E_{\mu''}$ instead of the summation $\sum_{\mu''}$ in the indirect-pathway term,
the main differences lie in the addition of an $E_p$ term in all exponents
and the introduction of the $R$-dependent weighting function $v(R)$
in the overlap integrals between the nuclear eigenstates of the resonance state and the final state.
As transitions between other states are not mediated by configuration interaction $V$, their pertinent overlap integrals acquire no such weighting function.

Therefore, the analytical solutions of the working equation derived in Ref.~\cite{Riegel25}
for a Gaussian-shaped laser pulse of duration $T_X$ with slowly varying envelope considering only the photon absorption,
for which the electric field is given by
\begin{align}
	\label{eq:E_field}
	\begin{split}
		\mathcal{E}_X (t') &= -\frac{\mathrm{d}}{\mathrm{d}t'}
		\left[A_{0X} \cos(\Omega t') \times \frac{1}{\sigma \sqrt{2\pi}} \exp\left(-\frac{{t'}^2}{2 \sigma^2}\right) \right] \\
		&\approx
        -\frac{A_{0X}\Omega}{2i\, \sigma\sqrt{2\pi}}
		\exp\left[-\left(i\Omega t' + \frac{{t'}^2}{2\sigma^2}\right)\right]	\, ,
	\end{split}
\end{align}
can readily be adapted for Eq.~\eqref{eq:ampl_working_eq}:
The adapted direct-pathway term is given by (compare to Eq.~(43) in Ref.~\cite{Riegel25})
\begin{widetext}
\begin{align}   \label{eq:dir_ampl}
\begin{split}
 &- \frac{A_{0X} \Omega \braket{c|\mu|g}}{2} \braket{\chi_\mu|\chi_0} \,
   \exp\biggl[ -it(E_{\text{kin}} + E_{\text{fin}} + E_\mu + E_p) \biggr] \,
   \exp\biggl[ -\frac{\sigma^2}{2} (E_{\text{kin}} + E_{\text{fin}} + E_{\mu} + E_p -\Omega)^2 \biggr] \, \\
   & \qquad \times \operatorname{Re} \Biggl[  \erf
   \biggl( \frac{1}{\sigma\sqrt{2}} \Bigl( \frac{T_X}{2}
   + i\sigma^2 \bigl( E_{\text{kin}} + E_{\text{fin}} + E_\mu + E_p - \Omega \bigr)
   \Bigr) \biggr) \Biggr] \, ,
\end{split}
\end{align}
\end{widetext}
where $\operatorname{Re}$ denotes the real part and $\erf$ denotes the error function.
The common term for resonance and indirect ionization, omitting their respective prefactors and weights which are
$\frac{\braket{r|\mu|g}}{N_\lambda} \braket{c|\alpha_V|r} \braket{\chi_\mu | v(R) | \chi_\lambda} \braket{\chi_\lambda | \chi_0}$
for the resonance pathway (and then $\sum_{\lambda}$) and
$- \frac{i\pi \braket{c|\mu|g}}{N_\lambda} \left|\braket{c|\alpha_V|r}\right|^2 \braket{\chi_\mu | v(R) | \chi_\lambda} \braket{\chi_\lambda | v(R) | \chi_{\mu''}} \braket{\chi_{\mu''} | \chi_0}$
for the indirect pathway (and then $\sum_{\lambda} \int \mathrm{d}E_{\mu''}$),
is given by (compare to Eq.~(44) in Ref.~\cite{Riegel25})
\begin{widetext}
\begin{align}   \label{eq:res_ampl}
\begin{split}
 &- \frac{A_{0X} \, \Omega}
         {4 (E_r + E_\lambda - i\pi W_\lambda - E_{\text{kin}} - E_{\text{fin}} - E_\mu)} \,
   \exp\biggl[ -it(E_r + E_\lambda + E_p - i\pi W_\lambda) \biggr] \\
   &\qquad \times \exp\biggl[ -\frac{\sigma^2}{2} (E_r + E_\lambda + E_p -i\pi W_\lambda -\Omega)^2 \biggr] \,
   \biggl( \erf(\tau_\text{1,max}) - \erf(\tau_\text{1,min}) \biggr) \\
 &+ \frac{A_{0X} \, \Omega}
         {4 (E_r + E_\lambda -i\pi W_\lambda - E_{\text{kin}} - E_{\text{fin}} - E_\mu)} \,
   \exp\biggl[ -it(E_{\text{kin}} + E_{\text{fin}} + E_\mu + E_p) \biggr] \\
   &\qquad \times \exp\biggl[ -\frac{\sigma^2}{2} (E_{\text{kin}} + E_{\text{fin}} + E_\mu +E_p -\Omega)^2 \biggr] \,
   \biggl( \erf(\tau_\text{2,max}) - \erf(\tau_\text{2,min}) \biggr)
\end{split}
\end{align}
\end{widetext}
with the transformed integral limits being
$\tau_\text{1,max} = \frac{1}{\sigma\sqrt{2}}
 \Bigl( \frac{T_X}{2} - i\sigma^2(E_r + E_\lambda + E_p -i\pi W_\lambda - \Omega) \Bigr)$,
$\tau_\text{1,min} = -\frac{1}{\sigma\sqrt{2}}
 \Bigl( \frac{T_X}{2} + i\sigma^2(E_r + E_\lambda + E_p -i\pi W_\lambda - \Omega) \Bigr)$,
$\tau_\text{2,max} = \frac{1}{\sigma\sqrt{2}}
 \Bigl( \frac{T_X}{2} - i\sigma^2(E_{\text{kin}} + E_{\text{fin}} +E_\mu + E_p - \Omega) \Bigr)$ and
$\tau_\text{2,min} = -\frac{1}{\sigma\sqrt{2}}
 \Bigl( \frac{T_X}{2} + i\sigma^2(E_{\text{kin}} + E_{\text{fin}} + E_\mu + E_p - \Omega) \Bigr)$.\\

The total spectrum is then determined by inserting all contributions into Eq.~\eqref{eq:td_probability},
where the sum over all nuclear final states is replaced by an integral.
\newline

\subsection{Final states of the ICD process}
\label{subsec:final_states}
In the final state, as discussed above, the two positively charged ions repel each other, leading to a Coulomb explosion.
The final-state PEC is therefore dissociative and hosts no nuclear vibrational states, only scattering states.
Since their wavefunctions differ qualitatively from the bound-state wavefunctions and knowledge about them aids in the understanding of ICD spectra, we will discuss their most important properties here.

For a one-dimensional repulsive Coulomb potential
\begin{equation}	\label{eq:Coulomb_pot}
	V_\text{Coulomb}(R) = \frac{V_a}{R} \quad , \, V_a > 0
\end{equation}
in a system with reduced mass $m$,
which is to be expected for a Coulomb explosion,
the momentum is $k(E) = \sqrt{2 m E}$
and the energy-normalized regular
(going to zero as $R$ goes to zero)
eigenfunctions are \cite{Yost36,Seaton02,bk:LandauLifshitz3}
\begin{equation}	\label{eq:CoulombWF_E-norm}
    \chi_E(R) = \sqrt{\frac{2 m}{\pi k}} F_{L=0}\left(\eta=\frac{V_a m}{k}, \rho=kR\right)
\end{equation}
with the regular Coulomb wavefunction for $L=0$,
\begin{align}	\label{eq:CoulombWF_k-norm}
    F_{L=0}(\eta,\rho) = &\exp\left(-\frac{\pi \eta}{2}\right) \left| \Gamma(1+ i\eta) \right| \nonumber\\
    &\times \rho \exp(-i\rho) M(1-i\eta,2,2i\rho)  \, ,
\end{align}
where $M(\alpha,\beta,z)$ is the confluent hypergeometric function $_1F_1$,
also known as Kummer function of the first kind,
which solves $z\Psi''(z) + (\beta-z) \Psi'(z) - \alpha \Psi(z) = 0$
and can be defined as
$M(\alpha,\beta,z) = \sum_{n=0}^\infty
\frac{\Gamma(\alpha+n)}{\Gamma(\alpha)}
\frac{\Gamma(\beta)}{\Gamma(\beta+n)} \frac{z^n}{n!}$
\cite{bk:Abramowitz1972}.

The eigenfunctions in Eq.~\eqref{eq:CoulombWF_E-norm} are employed as nuclear eigenfunctions $\chi_\mu(R)$ of the electronic final state.
In the derivation of the equations for the transition amplitude in Sec.~\ref{subsec:derivation}, we assumed that those are energy normalized,
\begin{equation}	\label{eq:Emu_norm}
	\braket{\chi_{\mu}|\chi_{\mu'}} = \delta(E_{\mu}-E_{\mu'})	\, .
\end{equation}
The prefactor in Eq.~\eqref{eq:CoulombWF_E-norm} ensures this \cite{bk:LandauLifshitz3,bk:Tannor2007}:
\begin{equation}	\label{eq:E_norm} 
	\braket{\chi_E|\chi_{E'}} = \delta(E-E')	\, .
\end{equation}

\begin{figure}[!h]
 \centering
 \includegraphics[width=\columnwidth]{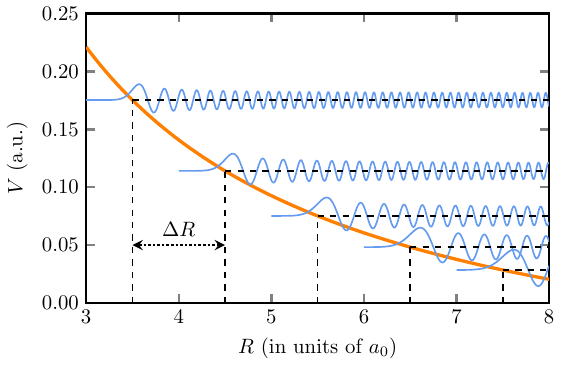}
 \caption{(Colour online)
 Repulsive Coulomb nuclear potential Eq.~\eqref{eq:Coulomb_pot} of the final state (orange) with a selection of
 energy-normalized nuclear Coulomb eigenfunctions Eq.~\eqref{eq:CoulombWF_E-norm} (blue).
 Energies are chosen for constant steps $\Delta R$ in the internuclear distance $R$ which correspond to non-uniformly distributed energies (dashed black lines) to better sample the region of interest.
 }
 \label{fig:final_wfs}
\end{figure}

The transition amplitude into a final state depends on [modified, where necessary, see Eq.~\eqref{eq:V_matrix_element}] Franck--Condon overlap integrals,
which in turn depend on the behaviour of the involved nuclear wavefunctions over the entire interval of internuclear distances $R$.
Some energy-normalized nuclear eigenfunctions of the electronic final state, Eq.~\eqref{eq:CoulombWF_E-norm}, are shown in Fig.~\ref{fig:final_wfs}.
They possess a maximum at around the classical turning point of the potential for the respective energy.
Inside the potential, the amplitude of the Coulomb wavefunction, Eq.~\eqref{eq:CoulombWF_k-norm}, decreases rapidly,
while the asymptotic behaviour for $\rho=kR\rightarrow\infty$ is
$F_{L=0}\rightarrow\sin\left[\rho - \eta \ln(2\rho) + \arg\Gamma(1+i\eta) \right]$ \cite{bk:LandauLifshitz3,bk:Abramowitz1972}.
The logarithmic term preventing the function
from asymptotically becoming a plane wave
is caused by the long-ranging Coulomb potential
(decreasing significantly slower than $\rho^{-2}$) \cite{bk:LandauLifshitz3,bk:Newton1982,bk:fano1986}.
For the bound--continuum Franck--Condon overlap integrals, however,
this asymptotic behaviour is unproblematic
because the amplitude of the bound-state vibrational wavefunctions of the electronic ground state or resonance state
decreases rapidly towards large distances.
The bound eigenfunctions of the Morse potential (see
Sec.~\ref{sec:CompDetails}) are an excellent example of this
exponential decrease in the limit of $R\to\infty$
(see Fig.~\ref{fig:energy_overview}).

In the numerical evaluation of Eq.~\eqref{eq:td_probability},
the integration over the energies $E_{\mu}$ of the nuclear final states
is replaced by a sum over discrete intervals.
If equal-sized energetic intervals $\Delta E$ are used, the integral can be approximated as
\begin{align}	\label{eq:P_DeltaE}
    P(E_\text{kin},E_p,t)
	&= \int_0^\infty \mathrm{d}E_\mu \left| \, \bra{E_\mu} \bra{E} \braket{p|\Psi(t)} \, \right|^2 \\
	&\approx \Delta E \times \sum_i \left| \, \bra{E_{\mu i}} \bra{E} \braket{p|\Psi(t)} \, \right|^2	\, .
\end{align}
With a repulsive Coulomb potential, Eq.~\eqref{eq:Coulomb_pot},
as the final-state potential,
a uniform distribution of points of integration or steps in the sum in $E_{\mu}$
would yield many points originating from the steep part of the hyperbolic potential, which the wavepacket rarely visits,
but only few from the shallow part, where most of the dynamics occur.
In order to achieve a more favourable distribution
of points upon numerical integration,
a variable change into the spatial domain can be performed via the inverse of Eq.~\eqref{eq:Coulomb_pot},
\begin{equation}	\label{eq:R_mu}
	R_{\mu} = \frac{V_a}{E_{\mu}}	\, ,
\end{equation}
see Fig.~\ref{fig:final_wfs}.
Denoting a nuclear final state explicitly by its energy as $\ket{E_\mu}$,
Eq.~\eqref{eq:P_DeltaE} then reads as follows:
\begin{equation}	\label{eq:P_Rmu}
    P(E_\text{kin},E_p,t) = \int_0^\infty
    \mathrm{d}R_\mu \left| \, \bra{E_\mu(R_\mu)} \bra{E} \braket{p|\Psi(t)} \, \right|^2 \rho^R(R_\mu)    \, .
\end{equation}
Here, $\rho^R(R_\mu)$ is the density of states with respect to $R_\mu$:
\begin{equation}	\label{eq:dos_R}
	\rho^R(R_\mu) = \rho^E(E_\mu(R_\mu)) \times \left| \frac{\mathrm{d}E_\mu}
	{\mathrm{d}R_\mu} \right|_{R_\mu}
	= 1 \times \frac{V_a}{R_\mu^2}
	= \frac{V_a}{R_\mu^2} \,.
\end{equation}
Discretizing the integral into a number of equal-sized finite $R$ intervals $\Delta R$ for numerical integration of Eq.~\eqref{eq:P_Rmu} yields:
\begin{equation}	\label{eq:P_DeltaR}
    P(E_\text{kin},E_p,t) \approx \Delta R \times
    \sum_i \left| \, \bra{E_\mu(R_i)} \bra{E} \braket{p|\Psi(t)} \, \right|^2  \rho^R(R_i)    \, .
\end{equation}

In practice, we limit the number of considered final nuclear states for computational reasons.
States above a threshold energy with negligible intensity of the laser pulse are not considered for the direct pathway and the first step of the indirect pathway.
For the resonance pathway, this is irrelevant; however, the physical limitation that the ICD must be energy-allowed for this pathway to be open restricts the high-energy final states in that case. In other words, a resonance state cannot decay into final states that lie energetically above the resonance state.
Finally, states for which the overlap integrals (with weighting function if necessary) with other vibrational states fall below a threshold are discarded as well.

\subsection{Wavepacket in the resonance state}
\label{subsec:wavepacket_res}

In order to analyze the decay process and its associated spectrum,
it may be useful to calculate the movement of the wavepacket in the resonance state.
This is possible in a fashion similar to the above derivation and the derivation in Ref.~\cite{Riegel25},
projecting the time-dependent wavefunction on a vibronic resonance state
instead of a final vibronic state [compare to the rear half of the second line of Eq.~\eqref{eq:trs_amplitude}]:
\begin{equation} \label{eq:wp_trs_amplitude}
		\bra{\chi_\lambda}\bra{r}\braket{p|\Psi(t)} = \,
		-i \int\limits_{t_0}^{t} \mathrm{d}t' \,
		\bra{\chi_\lambda}\bra{r}\braket{p|
			U_F(t,t') H_X(t') |G} \,.
\end{equation}
After some steps analogous to the ones above, we arrive at the following working equation for the resonance state
(analogous to Eq.~\eqref{eq:ampl_working_eq} for the final state):
\begin{widetext}
\begin{align}
	\begin{split}
		&\bra{\chi_\lambda}\bra{r}\braket{p|\Psi(t)} \\
		=& \, \frac{i}{N_\lambda} \braket{r|\mu|g}
		\braket{\chi_{\lambda} | \chi_0}
		\int\limits_{t_0}^{t} \mathrm{d}t' \,
		\exp[-i(E_r+E_{\lambda}+E_p-i \pi W_{\lambda})(t-t')]
		\mathcal{E}_X (t') \\
		&+ \frac{\pi}{N_\lambda}
		\braket{c|\mu|g} \braket{c|\alpha_V|r}^* \int \mathrm{d}E_{\mu''}
		\braket{\chi_{\lambda}| v(R) | \chi_{\mu''}}
		\braket{\chi_{\mu''} | \chi_0}
		\int\limits_{t_0}^{t} \mathrm{d}t' \,
		\exp[-i(E_r+E_{\lambda}+E_p-i \pi W_{\lambda})(t-t')]
		\mathcal{E}_X (t') \, .
	\label{eq:wp_ampl_working_eq}
	\end{split}
\end{align}
With the approximate electric field for the laser pulse,
specified in the second line of Eq.~\eqref{eq:E_field},
the analytical solution of Eq.~\eqref{eq:wp_ampl_working_eq} is given by
\begin{align}
	\begin{split}
 		& \left(
 		\frac{i}{N_\lambda} \braket{r|\mu|g}
 		\braket{\chi_{\lambda} | \chi_0}
 		+ \frac{\pi}{N_\lambda}
		\braket{c|\mu|g} \braket{c|\alpha_V|r}^* \int \mathrm{d}E_{\mu''}
		\braket{\chi_{\lambda}| v(R) | \chi_{\mu''}}
		\braket{\chi_{\mu''} | \chi_0}
 		\right) \\
 		& \quad \times \frac{-A_{0X}\Omega}{4i}
 		\exp\left[-it(E_r+E_{\lambda}+E_p-i \pi W_{\lambda})\right] \,
 		\exp\left[-\frac{\sigma^2}{2}(E_r+E_{\lambda}+E_p-i \pi W_{\lambda} -\Omega)^2\right] \\
 		& \quad \times
 		\left(
 		\erf\left(\tau_{\text{max}} \right)
 		-\erf\left(\tau_{\text{min}}\right)
 		 \right)
 		\label{eq:wp_analyt_easy}
 	\end{split}
\end{align}
with $\tau_{\text{max}} = \frac{1}{\sigma\sqrt{2}} \left(\frac{T_X}{2} - i\sigma^2 (E_r+E_{\lambda}+E_p-i\pi W_{\lambda} -\Omega) \right)$
and $\tau_{\text{min}} = -\frac{1}{\sigma\sqrt{2}} \left(\frac{T_X}{2} + i\sigma^2 (E_r+E_{\lambda}+E_p-i\pi W_{\lambda} -\Omega) \right)$.
\end{widetext}
The total wavepacket in the electronic resonance state is then constructed
by weighting the vibrational eigenfunctions with the transition amplitudes [Eqs.~\eqref{eq:wp_ampl_working_eq} or \eqref{eq:wp_analyt_easy}]
and adding up these wavefunctions as a superposition.

\section{Computational Details}
\label{sec:CompDetails}

\subsection{System and laser parameters}
\label{subsec:params}
We investigate the time-resolved ICD electron kinetic-energy spectrum in the neon dimer
involving the $1 ^{1}\Sigma_\mathrm{g}^+$ state of the neutral \ce{Ne-Ne} (ground state) \cite{Schnorr15,Stoychev08_2},
the $2 ^{2}\Sigma_\mathrm{u}^+$ state of the singly ionized \ce{Ne+({2s}^{-1})-Ne} (resonance state) \cite{Schnorr15,Stoychev08_2}
and one of the closely lying $^{1,3}\Sigma,\Pi,\Delta_{\text{g,u}}$ states of the doubly ionized \ce{Ne+({2p}^{-1})-Ne+({2p}^{-1})} (final state), namely the one used in Refs.~\cite{Scheit05,Scheit04}.

\begin{figure}[ht]
 \centering
 \includegraphics[width=0.74\columnwidth]{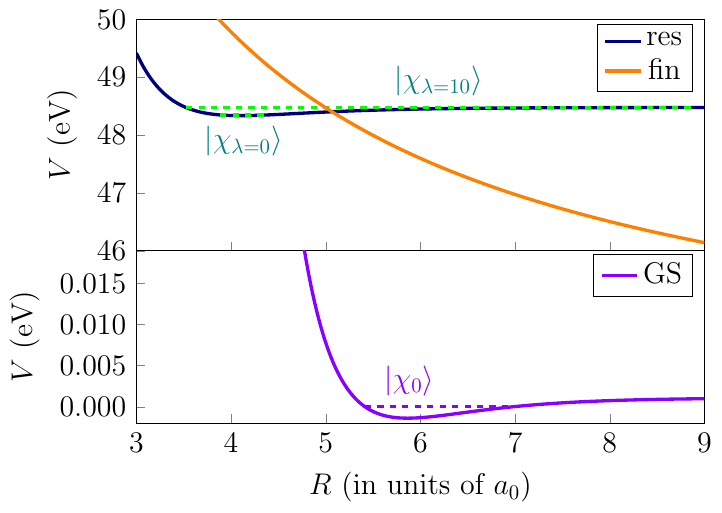}
 \caption{(Colour online) The PECs used to model the neon dimer.
 Top: PECs for the resonance state (blue curve),
 and for the final state (orange curve).
 The resonance state hosts 11 bound vibrational states $\ket{\chi_\lambda}$;
 the lowest and highest of their energy levels are indicated between their respective classical turning points (green dashed lines).
 Bottom: PEC for the electronic ground state (violet curve).
 The energetic level of the vibrational ground state $\ket{\chi_0}$ is indicated between the classical turning points (violet dashed line) and serves as the reference point for the energy scale.}
 \label{fig:Ne2_PEC}
\end{figure}

\begin{table}[h]
\centering
\caption{System and laser parameters used in the simulations of spectra.}
\label{tab:params}
$\begin{array}{rr}
\toprule
\multicolumn{2}{l}{\text{Ground state}} \\
D				& \qty{2.47}{\meV} \\
\alpha			& \qty{1.2434}{\per\bohr} \\
R_0				& \qty{5.86}{\bohr} \\
V_\text{min}	& \qty{-1.40}{\meV} \\
\midrule
\multicolumn{2}{l}{\text{Resonance state}} \\
D				& \qty{0.14}{\eV} \\
\alpha			& \qty{1.2106}{\per\bohr} \\
R_0				& \qty{4.10}{\bohr} \\
V_\text{min}	& \qty{48.34}{\eV} \\
\midrule
\multicolumn{2}{l}{\text{Final state}} \\
V_a				& \qty{26.2193}{\eV\bohr} \\
V_\text{min}	& \qty{43.23}{\eV} \\
\Delta R		& \qty{0.0189}{\bohr} \\
\midrule
\multicolumn{2}{l}{\text{Transition}} \\
v(R)			& R^{-3} \\
\tau(R_{0,\text{res}})	& \qty{82}{\fs} \\
q				& \num{100} \\
\midrule
\multicolumn{2}{l}{\text{Laser}} \\
I				& \qty{1}{\tera\W\per\square\cm} \\
\Omega			& \qty{58.2}{\eV} \\
n_{X,\,\text{short}}		& \num{60} \\
n_{X,\,\text{long}}		& \num{597} \\
\bottomrule
\end{array}$
\end{table}

The potential-energy curves (PECs) of these states were extracted through Ref.~\cite{WebPlotDigitizer} and fitted
using the trust-region-reflective least-squares algorithm \cite{Branch1999}
as implemented in the function \textsc{curve\_fit} in the module \textsc{scipy.optimize} \cite{SciPy-NMeth}.
The fitting functions were the Morse potential
\begin{equation}	\label{eq:Morse_pot}
	V_\text{Morse} = D \left\{1-\exp[-\alpha(R-R_0)]\right\}^2
\end{equation}
for the ground-state and the resonance-state PECs
and the repulsive Coulomb potential Eq.~\eqref{eq:Coulomb_pot} for the final-state PEC. This allows us to use analytic nuclear eigenfunctions
given by Refs.~\cite{Morse29,Scholz32,Pekeris34,Dahl88,Riegel25} for the Morse potentials and
by Eq.~\eqref{eq:CoulombWF_E-norm} for the final-state PEC.

The resulting PECs are shown in Fig.~\ref{fig:Ne2_PEC} and the fitting parameters are summarized in Table~\ref{tab:params}.
The minimum potential values $V_{\text{min}}$ reported therein are the minima of the Morse potentials and the asymptotic energy of the Coulomb potential, respectively.
They are defined relative to the vibronic ground state energy.

Solving the time-independent Schrödinger equation with these Morse potentials analytically yields
one bound vibrational state hosted by the electronic ground state and eleven bound vibrational states hosted by the electronic resonance state.
In reality, the ground state of the neon dimer encompasses two vibrational states.
Under the experimental conditions of Ref.~\cite{Schnorr15}, however, the vibrational ground state is populated to more than \qty{99}{\percent}. We therefore consider our model potential
suitable for a qualitative description of the time-dependent ICD electron spectra.


All Franck--Condon overlap integrals are calculated numerically from the known nuclear wavefunctions.
They are modified by a weighting function where necessary, see Eq.~\eqref{eq:V_matrix_element}.
For this, using the notation introduced in Eq.~\eqref{eq:V_of_R},
we choose $v(R) = R^{-3}$ and $\alpha_V = \text{const.} \in \mathbb{R}^+$ for all resonance and final states
so that at the internuclear distance corresponding to the minimum of the resonance-state Morse potential, $R_{0,\text{res}}$,
the lifetime is $\tau(R_{0,\text{res}}) = \frac1{\Gamma(R_{0,\text{res}})} = \frac1{2\pi \left|V(R_{0,\text{res}})\right|^2} = \qty{82}{\femto\second}$.
This is sensible because the distance-dependent part of the coupling operator,
$v(R)$ [see Eq.~\eqref{eq:V_of_R}], is proportional to $R^{-3}$
within the asymptotic approximation for the (dipole-allowed) ICD \cite{Averbukh04,Gokhberg10_1,Fasshauer13}.
The value of $\tau(R_{0,\text{res}})$ has been chosen in agreement with Refs.~\cite{Schnorr15,Averbukh06_1}.

For our simulations, we set the (electronic) Fano parameter to $q = 100$. Our reasoning is as follows:
For small systems and low to medium photon energies as those used in the experiment in Ref.~\cite{Schnorr15},
the probability of single ionization as required for the resonance pathway is considerably higher than the probability of direct single-photon double ionization into the final state as required by the other pathways.
Usually the ratio of double- to single-ionization atomic cross-sections for single-photon ionization is approximately less than one percent or up to only a few percent for the energies we are concerned with \cite{Bartlett92,Dorner04,Bluett05,Kheifets07}.
Therefore, we expect the Fano parameter $q$, which quantifies the relative contributions to the transition amplitude via the resonance pathway and the direct and the indirect pathways \cite{Fano61},
to be significantly larger than unity.
Here we have chosen $q = 100$, which virtually excludes all pathways but the resonance one.

The ionizing XUV laser pulse is largely modelled after the one used in the experiments of Ref.~\cite{Schnorr15}:
an intensity of \qty{1e12}{\watt\per\square\centi\metre}
and a central photon energy of $\Omega = \qty{58.2}{\eV}$.
Two different pulse durations have been chosen:
an FWHM of the Gaussian envelope of the electric field of \qty{6.0}{\fs} corresponding to $n_X = 60$ cycles within the FWHM for a shorter pulse,
and an FWHM of \qty{60}{\fs} corresponding to $n_X = 597$ cycles and $\sigma = \qty{26}{\meV}$ for a longer pulse.
The total pulse width used in the simulations is defined as $5\sigma$ in both cases.
In the experiment, the central frequency of the free-electron laser fluctuated by roughly \qty{1}{\eV}.
We have performed simulations with different central frequencies sampled from this interval and found only minor changes to the spectra.
Therefore, we limit ourselves to only the above value for the central frequency.
\\

All simulations were performed using the script \textsc{nuclear\_dyn} within our \textsc{eldest} programme package \cite{ELDEST_v3.0}.
Input files can be found in Ref.~\cite{figshare_input_ICD_2026}.
\\

\subsection{Discretization of the final-state nuclear continuum}
\label{subsec:discrete}

For the technical reasons discussed in Sec.~\ref{subsec:final_states}, we discretize the continuum of final nuclear eigenstates $\{\ket{\chi_\mu}\}$ in constant $R$ steps.
This will have an effect on the simulated spectra which we want to address here.

\begin{figure}[!ht]
 \centering
 \includegraphics[width=\columnwidth]{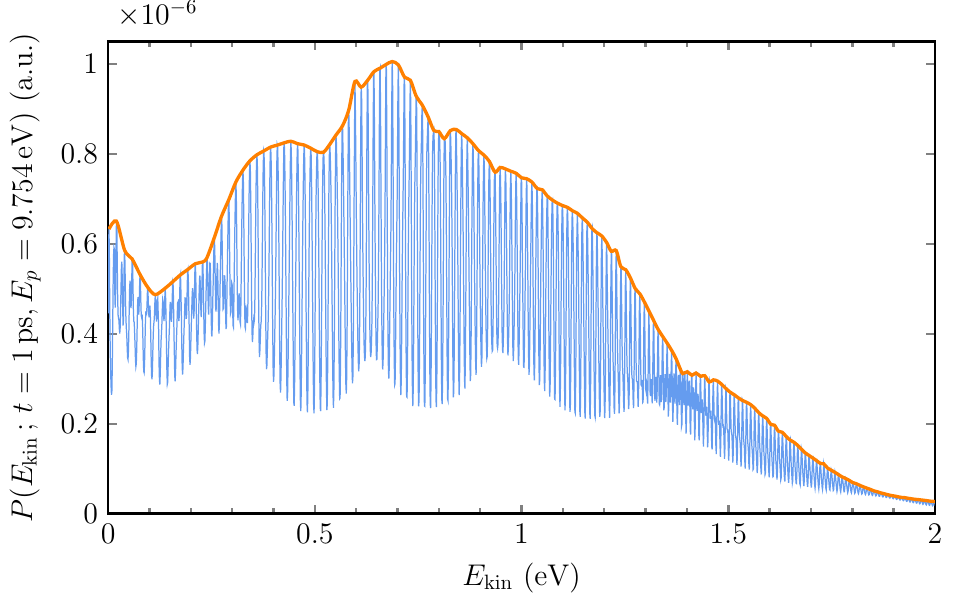}
 \caption{(Colour online) ICD electron spectrum (blue curve)
  as function of the kinetic energy of the secondary electron, $E_\text{kin}$,
 at constant time $t = \qty{1}{\ps}$ and kinetic energy of the photoelectron, $E_p = \qty{9.754}{\eV}$. The ionizing laser pulse has FWHM of \qty{6}{\fs}.
 The overall structure is physical (orange curve on top, see Sec.~\ref{subsec:ICD_electron}),
 the fine structure is an artefact due to the discretization of the continuum of nuclear final states in steps of \qty{0.01}{\AA}, see main text.
 }
 \label{fig:spec_t1000fs_fwhm6fs}
\end{figure}

A typical ICD electron spectrum is shown in Fig.~\ref{fig:spec_t1000fs_fwhm6fs}.
For this simulation, a constant discretization step width of $\Delta R = \qty{0.0189}{\bohr} = \qty{0.01}{\AA}$ has been used, see Eq.~\eqref{eq:P_DeltaR}.
In addition to the broader overall structure, the simulated spectrum possesses
an underlying fine structure with spacings decreasing from \qty{20}{\milli\eV} for low values of $E_\text{kin}$ to \qty{8}{\milli\eV} for high values of $E_\text{kin}$.
This fine structure is absent in similar spectra reported in literature for the neon dimer \cite{Jahnke04,Jahnke07,Higuchi10,Kim14,Yan18}.
We find, in fact, that it is an artefact arising from the discretization of the continuum of nuclear final states:

The spacing of the rapidly oscillating peaks matches the discretization steps.
It follows from Eq.~\eqref{eq:R_mu} that the constant spatial discretization step width of $\Delta R = \qty{0.01}{\AA}$ translates to energetic step widths
$\left|{\Delta E}_i\right| = \left|E_\mu(R_i + \Delta R)-E_\mu(R_i)\right| = \frac{V_a \Delta R}{R_i(R_i + \Delta R)} \approx \frac{V_a \Delta R}{R_i^2}$.
This yields step widths of \qty{20}{\milli\eV} near the intersection of the PECs of resonance state and final state (corresponding to $E_\text{kin} = \qty{0}{\eV}$)
to \qty{8}{\milli\eV} at around $E_\text{kin} = \qty{2}{\eV}$.
These energetic steps agree with the fine spacings found in the simulated spectrum.

Furthermore, simulations with a spatial discretization step twice as large lead to a similar spectrum but with the fine energetic spacings approximately twice as large.
This confirms that the narrow peaks correspond to different discrete nuclear final states resulting from the discretization in our concrete implementation.

In the limit of infinitely fine resolution, that is, the true continuum of nuclear final states, the artificial substructure should vanish and only the overall structure remain.
We can, therefore, disregard the rapid oscillations for interpretation of the spectrum (Sec.~\ref{subsec:ICD_electron})
and focus on the curve formed by the maxima
as indicated in Fig.~\ref{fig:spec_t1000fs_fwhm6fs}. \\

%

For a quantitative measure of the quality of our results with respect to the effects of the discretization, we checked if
the analogue of the Franck--Condon factor sum rule for a continuum of nuclear final states
is fulfilled,
\begin{equation}	\label{eq:fc_sumrule_int}
    \int \mathrm{d}E_\mu \, \left| \braket{\chi_n|\chi_\mu} \right|^2 = 1
    \,  ,
\end{equation}
or, following Eqs.~\eqref{eq:P_Rmu} to \eqref{eq:P_DeltaR}, for equal-sized spatial intervals,
\begin{equation}	\label{eq:fc_sumrule_deltaR}
	\Delta R \times \sum_i \left| \braket{\chi_n|\chi_{\mu i}} \right|^2
	\rho^R(R_i) \approx 1  \,  .
\end{equation}
Here, $\ket{\chi_n}$ represents any discrete vibrational eigenstate not hosted by the electronic final state,
that is, $\ket{\chi_0}$ of the ground state or a $\ket{\chi_{\lambda}}$ of the resonance state.
Since the quantity $\left| \braket{\chi_n|\chi_\mu} \right|^2$ has the dimension of reciprocal energy,
it is more suitable to call it a Franck--Condon density instead of a Franck--Condon factor (see, for example, Footnote~16 in Ref.~\cite{Gislason73}).

We consider the discretization resolution sufficiently fine if the left-hand-side quantity in Eq.~\eqref{eq:fc_sumrule_deltaR}
deviates from unity by no more than \num{1e-5}.
That condition is fulfilled for a step width of $\Delta R = \qty{0.01}{\AA}$,
for which the deviation from unity is less than \num{2e-6} for all bound states.
Further numerical testing indicated that with this resolution,
the decay lifetimes extracted from the simulated time-dependent spectra (Sec.~\ref{subsec:lifetime}) can be considered converged as well.

Therefore, constant steps of $\Delta R = \qty{0.0189}{\bohr} = \qty{0.01}{\AA}$
have been used in all simulations discussed in Sec.~\ref{sec:results}.
For the laser pulse with FWHM of \qty{6}{\fs}, this corresponds to 670 states
of which up to 494 can be reached via ICD from one of the 11 bound vibronic resonance states;
for the laser pulse with FWHM of \qty{60}{\fs}, for which the maximum photon energy is smaller,
it corresponds to 666 states of which up to 493 can be reached via ICD.
\section{Results}
\label{sec:results}

The results are presented as follows:
We begin in Sec.~\ref{subsec:photoelectron} with a short discussion of how we choose the value of the photoelectron energy that we fix in our subsequent simulations.
Then we analyze the ICD electron spectrum at large times in Sec.~\ref{subsec:ICD_electron}.
We particularly focus on the extent to which the spectrum can or cannot be explained in terms of Franck--Condon factors and densities without interference terms.
In Sec.~\ref{subsec:td_spec}, we address the temporal evolution of the ICD spectrum,
which can be rationalized in terms of the movement of the decaying wavepacket in the electronic resonance state.
Finally in Sec.~\ref{subsec:lifetime}, we demonstrate how the decay lifetime can be extracted from the time-resolved decay spectra,
both for simulated and experimental ones.
We establish that for a correct interpretation, the time-resolved integral spectral intensity must be fitted to a double-exponential function of time as follows from our analytical expressions.

\subsection{Central photoelectron energy}
\label{subsec:photoelectron}

\begin{figure}[ht]
 \centering
 \includegraphics[width=\columnwidth]{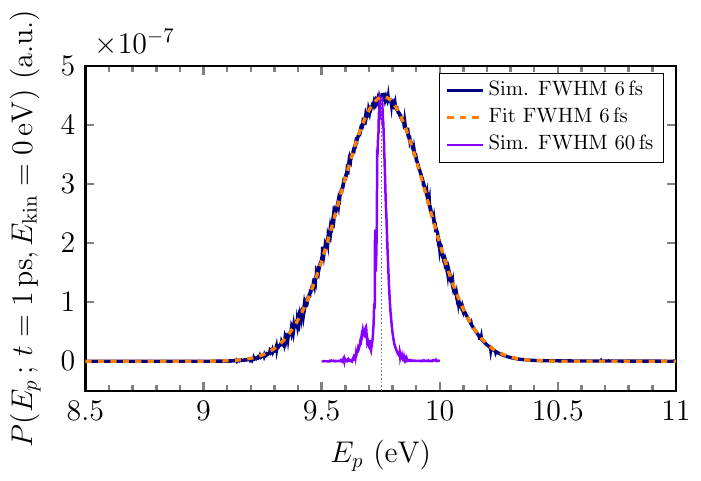}
 \caption{(Colour online) ICD signal intensity (dark blue solid line) as a function of the kinetic energy of the photoelectron, $E_p$, at constant time $t = \qty{1}{\ps}$ and kinetic energy of the ICD electron, $E_\text{kin} = \qty{0}{\eV}$. The ionizing laser pulse has FWHM of \qty{6}{\fs}. The spectral shape can be closely fitted to a Gaussian function (orange dashed line). Spectrum for the pulse with FWHM of \qty{60}{\fs} shown for comparison (violet solid line), a few data points affected by numerical noise have been discarded and the spectrum has been scaled up. Central photoelectron energy used in subsequent simulations is indicated.}
 \label{fig:Ep_sweep_both}
\end{figure}

In our simulations of time-resolved ICD electron spectra,
we want to limit the photoelectron energy to a single value of high intensity in the interest of reduced computational cost.
For the purpose of selecting this energy,
we calculate the double-ionization probability, Eq.~\eqref{eq:td_probability}, as a function of the photoelectron energy $E_p$
for a constant kinetic energy $E_\text{kin} = \qty{0}{\eV}$ of the secondary ICD electron.
The result is shown in Fig.~\ref{fig:Ep_sweep_both} for $t = \qty{1}{\ps}$.

The photoelectron spectrum for a laser pulse FWHM of \qty{6}{\fs} 
can be described well by a Gaussian distribution.
Fitting the spectrum yields a central energy of \qty{9.761}{\eV}.
The observed shape is consistent with the Gaussian-shaped ionizing XUV laser pulse,
which is centred on the vibrational ground state of the electronic resonance state and coherently populates the different vibrational states.

We obtain similar results for a pulse with FWHM of \qty{60}{\fs},
albeit with a narrower spectral width and therefore less population in the higher-lying vibrational states.
The central photoelectron energy for this longer pulse is \qty{9.754}{\eV}.

Energies close to these central values are sensible choices for the fixed photoelectron energy in subsequent simulations.
The spectrum for the 60-fs pulse is narrower, so that deviations from the maximum lead to a more severe loss of signal intensity.
Therefore, we choose the central value for the 60-fs pulse as fixed photoelectron energy, $E_p = \qty{9.754}{\eV}$.
This is also close enough to the 6-fs maximum (see the dashed vertical line in Fig.~\ref{fig:Ep_sweep_both}).
\\
\subsection{ICD electron spectrum}
\label{subsec:ICD_electron}

\begin{figure*}[t]
 \centering
 \includegraphics[width=\columnwidth]{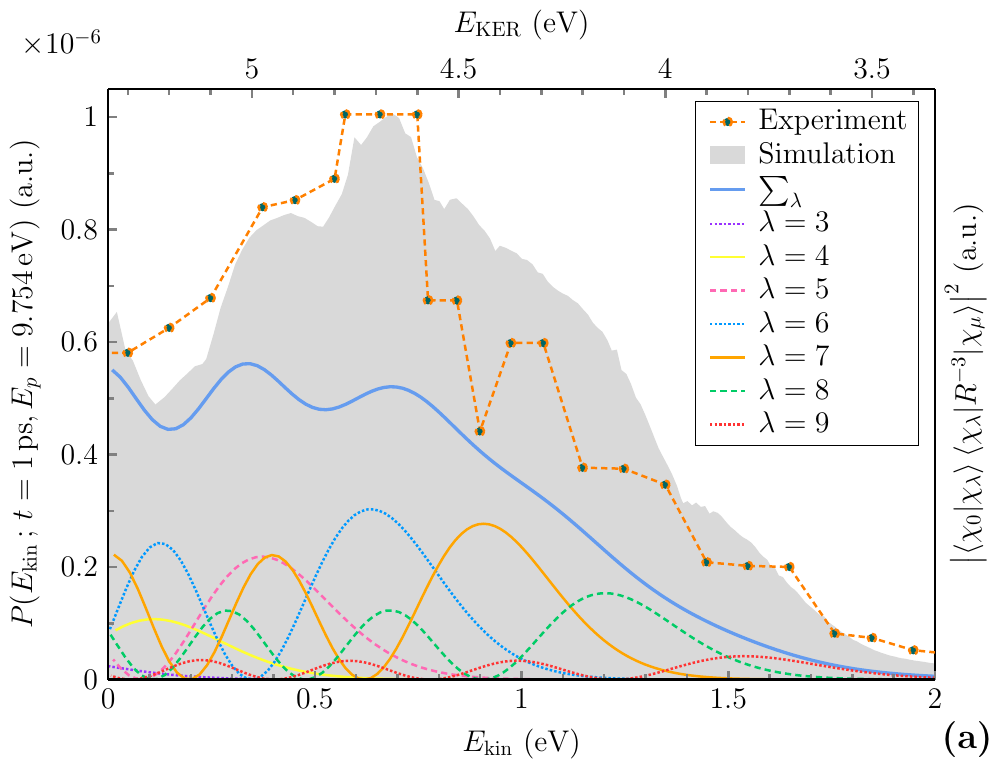}
 \hfill
 \includegraphics[width=\columnwidth]{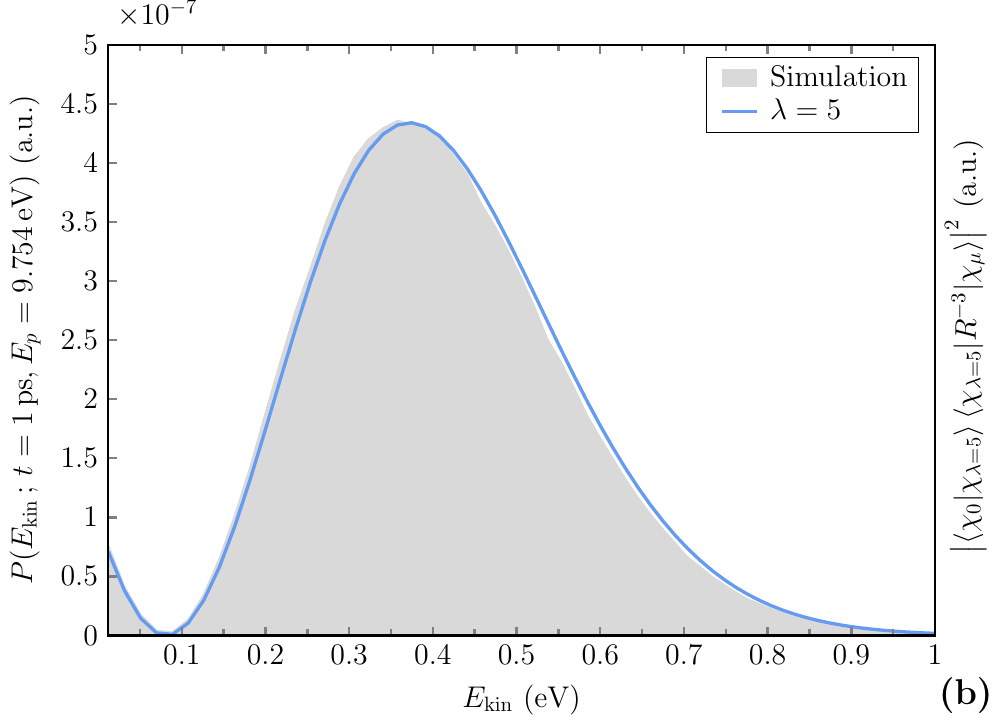}
 \caption{(Colour online)
 Simulated ICD spectrum as a function of the kinetic energy of the secondary electron, $E_\text{kin}$,
 at constant time $t = \qty{1}{\ps}$ and kinetic energy of the photoelectron, $E_p = \qty{9.754}{\eV}$ (grey area), where the ionizing laser pulse has FWHM of \qty{6}{\fs}.
 Weighted Franck--Condon densities $\left|\braket{\chi_0|\chi_{\lambda}}\braket{\chi_{\lambda}|R^{-3}|\chi_{\mu}}\right|^2$ (coloured curves) overlaid for the dominant vibrational resonance states, where $\ket{\chi_{\mu}}$ is the nuclear final state with energy $E_\mu$.
 These densities are plotted against $E_\text{kin} = E_r + E_\lambda - E_\text{fin} - E_\mu$ so that the energy axis of the Franck--Condon curves is consistent with the energy axis of the spectrum.
 (a) All bound vibrational resonance states are used in the simulation. Sum of the partial Franck--Condon curves shown as well (uppermost solid blue curve).
 For comparison, experimental spectrum (orange-green markers, orange dashed line meant to guide the eye) from Ref.~\cite{Schnorr15}.
 The latter was recorded as a KER spectrum, which we have converted into a kinetic-energy spectrum for the ICD electron,
 shifted by \qty{0.1}{\eV} for best agreement with the simulated spectrum.
 (b) Only the vibrational resonance state $\ket{\chi_{\lambda=5}}$ is considered in the simulation.
}
 \label{fig:spec_fc}
\end{figure*}

We now turn to the kinetic-energy spectrum of the ICD electron and
compare our theoretical results with experiment.
We consider the simulated ICD electron spectrum at a fixed time
$t = \qty{1}{\ps}$ and constant photoelectron energy
$E_p = \qty{9.754}{\eV}$.
The experimental neon dimer spectrum is derived from the kinetic energy
release (KER) spectrum reported in Ref.~\cite{Schnorr15}, which we
have converted into the corresponding kinetic-energy spectrum of the
ICD electron.
In Fig.~\ref{fig:spec_fc}(a) we compare it to our simulated spectrum
for a 6-fs laser pulse, previously discussed in
Sec.~\ref{subsec:discrete} and shown in
Fig.~\ref{fig:spec_t1000fs_fwhm6fs}.

Despite the markedly different pulse durations of \qty{6}{\fs} in the
simulation and \qty{60}{\fs} in the experiment, the theoretical and
experimental spectra exhibit a remarkably similar overall structure,
with multiple broad peaks at energies ranging from \qty{0}{\eV} to
\qty{2}{\eV}.
$E_{\text{kin}} = \qty{0}{\eV}$ corresponds to the crossing of the
vibronic energy levels of the resonance and final states, where the
ICD channel opens.
We note that this range up to circa \qty{2}{\eV}, matching many
experimental results for the neon dimer
\cite{Jahnke04,Jahnke07,Higuchi10,Kim14,Yan18}, is well separated
from the energy range of the photoelectron, which starts at circa
\qty{9}{\eV}.
The two electrons can, therefore, be clearly distinguished by their
kinetic energies in coincidence measurements.

We begin our analysis with the 6-fs case, as it provides a useful
starting point for understanding the origin of the spectral structure
before turning to the experimentally relevant pulse duration.

\subsubsection{Franck--Condon analysis}
\label{subsubsec:ICD_FC}
The overall structure of the spectrum can to a large degree be explained by deconstructing it into the contributions of the individual vibronic resonance states. 
This is shown in Fig.~{\ref{fig:spec_fc}(a)}, where the sum of weighted resonance state--final state Franck--Condon densities as well as the individual contributions of the dominant vibrational resonance states are shown.

The contribution of each vibrational state to the shape of the spectrum is determined within this model
by mapping the nodal structure of the vibrational wavefunctions in the resonance state
onto the final-state PEC \cite{moiseyev2001,Scheit03,Sisourat10_1}.
This mapping is usually performed in the literature via the reflection approximation \cite{Condon1928,Gislason73,Weber04}.
For our system, it is not necessary to invoke the reflection approximation,
because we know the analytical nuclear wavefunctions for the relevant PECs.
We have, therefore, directly used the Franck--Condon densities,
including the weighting function $v(R) = R^{-3}$ (see Eq.~\eqref{eq:V_of_R}, as argued for in Ref.~\cite{Sisourat10_1}).
In Fig.~{\ref{fig:spec_fc}(a)}, the peaks of the spectrum can be matched to one or more peaks in the weighted Franck--Condon densities.
These in turn reflect the nodal structure of the vibrational states.
Beyond \qty{2}{\eV}, the weighted Franck--Condon densities are too small, so ICD becomes much less probable compared to slower ICD electrons.

\begin{table}[h]
\centering
\caption{Franck--Condon overlap integrals and factors
for the vibrational ground state of the electronic ground state, $\ket{\chi_0}$,
and the bound vibrational states of the electronic resonance state, $\ket{\chi_{\lambda}}$,
as well as their energy relative to the minimum of the resonance-state PEC.}
\label{tab:gs-res-fcs}
\begin{tabular}{rrrr}
\toprule
$\lambda$ & $\braket{\chi_{\lambda} | \chi_0}$ &
     $\left|\braket{\chi_{\lambda} | \chi_0}\right|^2$ & $E_\lambda (\unit{\eV})$ \\
\midrule
 0 &  0.003  & \num{8e-06}	& 0.012	\\
 1 &  0.014  & \num{2e-04}	& 0.035	\\
 2 &  0.043  & 0.002		& 0.055	\\
 3 &  0.106  & 0.011		& 0.073	\\
 4 &  0.209  & 0.044		& 0.089	\\
 5 &  0.341  & 0.116		& 0.103	\\
 6 &  0.464  & 0.215		& 0.115	\\
 7 &  0.522  & 0.273		& 0.124	\\
 8 &  0.473  & 0.223		& 0.132	\\
 9 &  0.317  & 0.101		& 0.137	\\
10 &  0.119  & 0.014		& 0.140	\\
\bottomrule
\end{tabular}
\end{table}

As can be seen from the bottom part of Fig.~{\ref{fig:spec_fc}(a)}, the different resonance decay channels contribute to different extents.
The most prominent contributions arise from the vibrational resonance states $\ket{\chi_{\lambda=6}}$ and $\ket{\chi_{\lambda=7}}$,
with further appreciable contributions from $\ket{\chi_{\lambda=5}}$ and $\ket{\chi_{\lambda=8}}$.
This is determined by the ground state--resonance state Franck--Condon factors,
which are given in Table~\ref{tab:gs-res-fcs}.
Weighting the resonance state--final state Franck--Condon densities with these factors takes into account the probability of the laser pulse populating the respective vibrational level of the resonance state in the first place.
It is also in accordance with the resonance-pathway part of Eq.~\eqref{eq:ampl_working_eq} in conjunction with Eq.~\eqref{eq:td_probability}.
The above states are the ones with the largest ground state--resonance state Franck--Condon factors:
Because the minima of the ground-state and resonance-state PECs are displaced relative to each other by almost \qty{1}{\AA},
see Fig.~\ref{fig:Ne2_PEC},
the overlap integral between their respective vibrational ground states, $\braket{\chi_{\lambda=0} | \chi_0}$, is small,
while the overlap with higher lying states $\ket{\chi_{\lambda}}$ is considerably more favourable.
Therefore, there is a large probability that these are the initial states for ensuing ICD, hence their contribution to the sum spectrum being large.
Then, a large resonance state--final state Franck--Condon density for a certain energy, which corresponds to a certain final state, translates to a high probability of decay into this final state.

In principle, an additional weighting of the different vibrational states is introduced by the Gaussian spectral distribution of the ionizing laser pulse.
This means the individual state contributions should also be weighted by
a Gaussian distribution according to the relative energies of the vibrational states,
see Eq.~\eqref{eq:res_ampl}.
This plays an insignificant role for the laser pulse with FWHM of \qty{6}{\fs}:
Centred on the vibrational ground state of the resonance state,
it has a spectral FWHM of \qty{0.61}{\eV}.
Meanwhile, the 11 resonance states lie within \qty{0.13}{\eV} of each other, see Table~\ref{tab:gs-res-fcs}.
The short-pulse intensity for ionization into any of the vibrational resonance states is therefore similar.
Hence, this effect is mostly negligible for this pulse in comparison to the Franck--Condon factors, which vary by several orders of magnitude. \\

\subsubsection{Impact of vibronic resonance interferences}
\label{subsubsec:ICD_interfere}
Despite the overall agreement, discrepancies between the simple sum model and the full experimental and simulated spectra in Fig.~{\ref{fig:spec_fc}(a)} remain.
While the model largely reproduces the peak positions, it does
not fully account for their relative intensities, with the discrepancy
becoming particularly apparent at higher kinetic energies of the ICD
electron.

The origin of this discrepancy lies in the way contributions from
different vibronic resonance states combine in the total spectrum.
In the simple sum model, the contributions from individual vibronic
resonance states are represented by weighted Franck--Condon densities
and factors and subsequently summed.
In the full calculation, by contrast, the summation over all resonance
states in Eq.~\eqref{eq:ampl_working_eq} is performed \emph{prior} to
taking the absolute square in Eq.~\eqref{eq:td_probability}.
The simplified Franck--Condon analysis therefore neglects interference
terms between different vibronic resonance states.

To isolate the effect of these interference terms, we recompute the
ICD spectrum considering only the contribution from a single vibronic
resonance state, thereby eliminating all cross-terms between different
resonance states.
For each vibronic resonance state, the resulting spectrum agrees with
the corresponding weighted Franck--Condon contribution.
One representative example is shown in Fig.~\ref{fig:spec_fc}(b).
This confirms that the deviations of the full ICD spectrum from the
simple Franck--Condon picture arise from interference between different
vibronic resonance states.

\subsubsection{Dependence on pulse duration}
\label{subsubsec:ICD_60fs}

\begin{figure}[htb]
 \centering
 \includegraphics[width=\columnwidth]{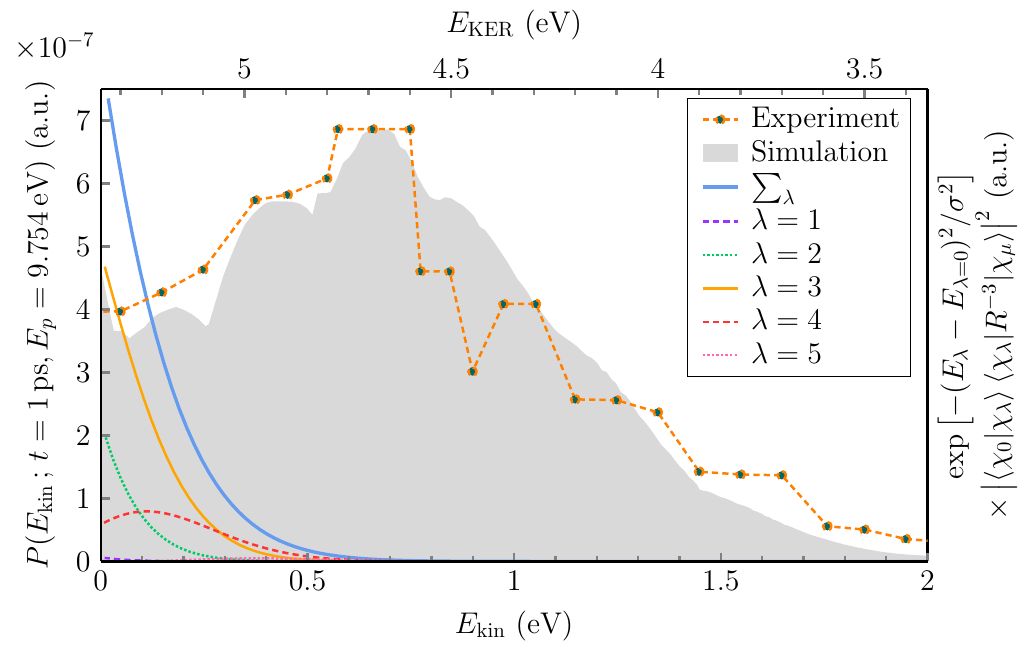}
 \caption{(Colour online) Simulated ICD spectrum as a function of the kinetic energy of the secondary electron, $E_\text{kin}$,
 at constant time $t = \qty{1}{\ps}$ and kinetic energy of the photoelectron, $E_p = \qty{9.754}{\eV}$ (grey area), where the ionizing laser pulse has FWHM of \qty{60}{\fs} corresponding to $\sigma = \qty{26}{\meV}$.
 Weighted Franck--Condon densities $\exp\left[-(E_{\lambda}-E_{\lambda=0})^2/\sigma^2\right] \left|\braket{\chi_0|\chi_{\lambda}}\braket{\chi_{\lambda}|R^{-3}|\chi_{\mu}}\right|^2$ (coloured curves) overlaid for the dominant vibrational resonance states, where $\ket{\chi_{\mu}}$ is the nuclear final state with energy $E_\mu$.
 These densities are plotted against $E_\text{kin} = E_r + E_\lambda - E_\text{fin} - E_\mu$ so that the energy axis of the Franck--Condon curves is consistent with the energy axis of the spectrum.
Sum of the partial Franck--Condon curves shown as well (uppermost solid blue curve). For comparison, experimental spectrum (orange-green markers, orange dashed line meant to guide the eye) from Ref.~\cite{Schnorr15}.
 The latter was recorded as a KER spectrum, which we have converted into a kinetic-energy spectrum for the ICD electron,
 shifted by \qty{0.1}{\eV} for best agreement with the simulated spectrum.
}
 \label{fig:spec_60fs_fc}
\end{figure}

We now turn to the longer ionizing pulse with a FWHM of
\qty{60}{\fs}, corresponding to the pulse duration employed in the
experiment \cite{Schnorr15}.
The resulting ICD electron spectrum is shown in
Fig.~\ref{fig:spec_60fs_fc}.
It is qualitatively similar to that obtained for the 6-fs pulse in
Fig.~\ref{fig:spec_fc}(a), with slightly better agreement with the
experimental spectrum, as may be expected for the pulse duration used
in the experiment.

Its decomposition into contributions from individual vibronic
resonance states, however, reveals a markedly different picture.
For the 6-fs pulse, the main features of the ICD electron spectrum
could still largely be understood from the sum of these individual
contributions.
For the 60-fs pulse, this picture breaks down.
The sum curve (upper blue curve in Fig.~\ref{fig:spec_60fs_fc})
accounts only for the low-energy part of the ICD spectrum up to
approximately \qty{0.5}{\eV} and does not reproduce the peak structure,
even at low kinetic energies.
Thus, while the full spectrum changes comparatively little upon
increasing the pulse duration, the role of interference in shaping the
spectrum changes fundamentally.

To understand this difference, we first consider how the spectral
width of the ionizing pulse affects the population of the vibronic
resonance states.
Increasing the temporal FWHM from \qty{6}{\fs} to \qty{60}{\fs}
reduces the spectral FWHM by a factor of ten to \qty{0.06}{\eV},
which is in this case comparable to the energetic spacings between the vibrational
resonance states, see Table~\ref{tab:gs-res-fcs}.
The pulse parameters therefore strongly favour the low-lying vibronic
resonance states, with the available pulse intensity decreasing
towards higher $\lambda$ and becoming very small for
$\lambda=5$, 6 and 7.

The ground state--resonance state Franck--Condon factors show a very
different distribution.
They particularly favour the states $\lambda=5$--8, with a maximum at
$\lambda=7$.
These states, however, receive little intensity from the spectrally
narrow ionizing pulse.
The individual contributions to the simple sum model are therefore
determined by the interplay between these two competing factors.

The low-lying states favoured by the pulse have comparatively small
Franck--Condon factors, whereas the states with the largest
Franck--Condon factors are only weakly populated by the pulse.
As a result, the largest individual contributions arise for the
intermediate states, in particular $\ket{\chi_{\lambda=3}}$ and
$\ket{\chi_{\lambda=4}}$, as shown in
Fig.~\ref{fig:spec_60fs_fc}.
Yet their sum cannot explain the full spectrum.
The corresponding weighted Franck--Condon densities are concentrated
at low kinetic energies and cannot account for the spectral intensity
extending beyond \qty{0.5}{\eV} or for the pronounced structure of the
full ICD electron spectrum.

The missing terms are the interferences between different vibronic
resonance pathways discussed in Sec.~\ref{subsubsec:ICD_interfere}.
In the full calculation, the sum over the vibrational resonance states
in Eq.~\eqref{eq:ampl_working_eq} is taken before the absolute square
in Eq.~\eqref{eq:td_probability}, giving rise to cross-terms that are
absent from the simple sum model.
For the 60-fs pulse, these cross-terms dominate the ICD electron
spectrum.
The spectrum can therefore no longer be understood primarily in terms
of the individual resonance-state contributions.
\subsection{Time-dependent ICD spectrum}
\label{subsec:td_spec}

\begin{figure}[!ht]
 \centering
 \includegraphics[width=\columnwidth]{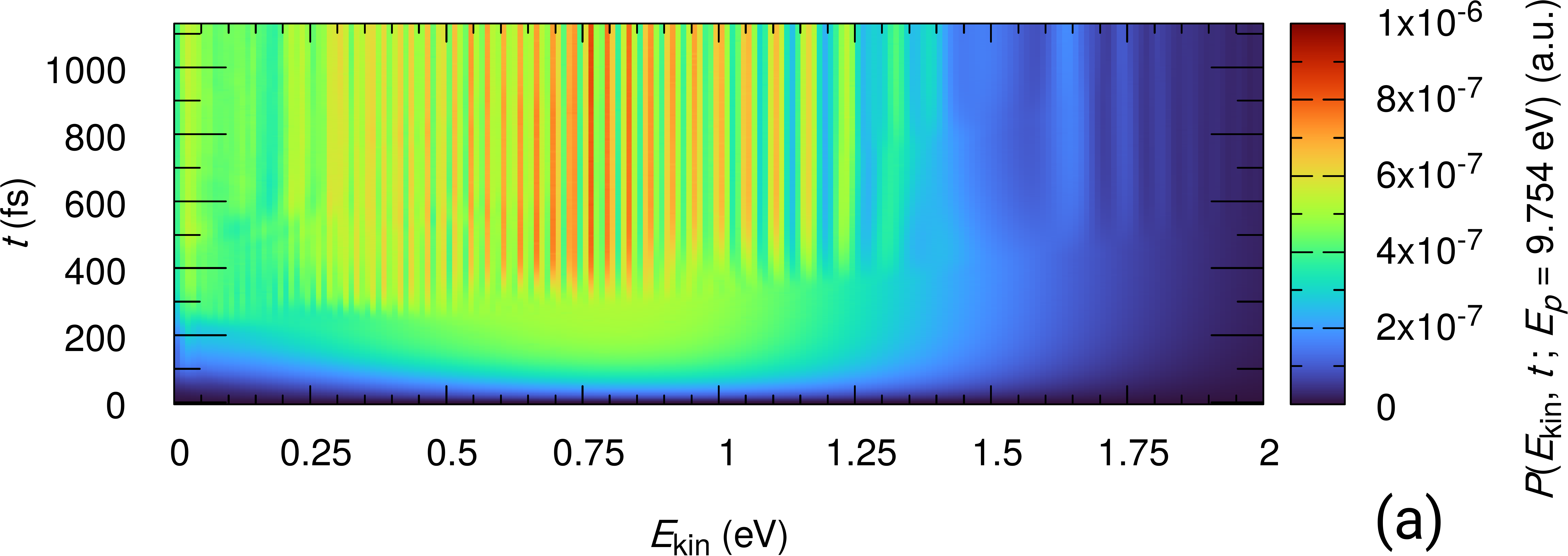}\\
 \vspace{0.5cm}
 \includegraphics[width=\columnwidth]{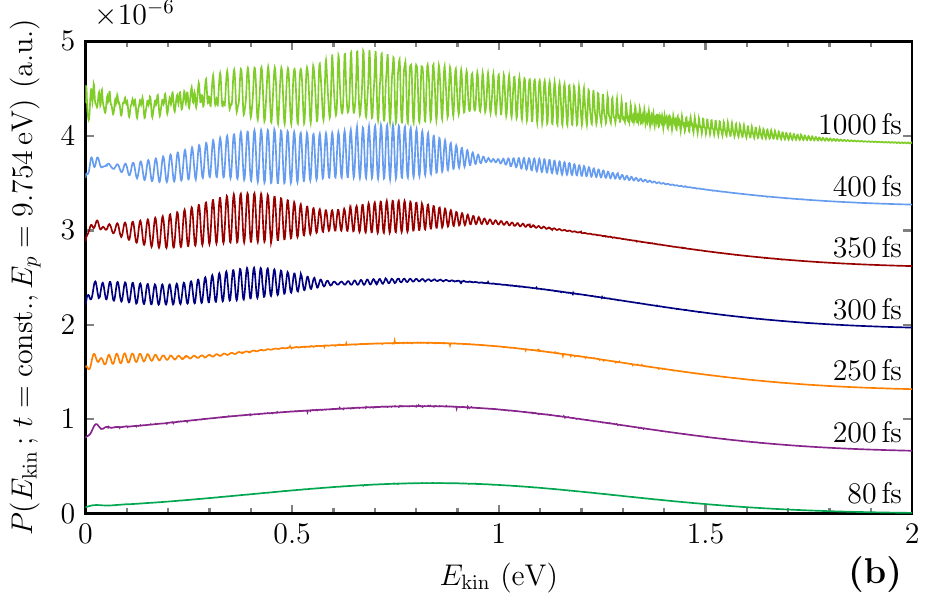}
 \caption{(Colour online) ICD electron spectrum as a function of the kinetic energy of the secondary electron, $E_\text{kin}$,
 and time, $t$, at constant kinetic energy of the photoelectron, $E_p = \qty{9.754}{\eV}$. The ionizing laser pulse has FWHM of \qty{6}{\fs}.
 (a) From-above map-like view, where colour indicates the ICD signal intensity.
 (b) Spectra at selected constant times after the beginning of the ionizing pulse, vertically displaced against each other for better visibility.}
 \label{fig:3d_spec_6fs}
\end{figure}

The time evolution of the ICD electron spectrum is shown in
Fig.~\ref{fig:3d_spec_6fs} for the short 6-fs pulse \cite{link:suppl_lifetime_2026}.
Figure~{\ref{fig:3d_spec_6fs}(a)} displays the full time- and
energy-resolved spectrum.
The rapid oscillations along the energy axis are discretization
artifacts resulting from the treatment of the final-state continuum,
as discussed in Sec.~\ref{subsec:discrete}.
To examine the development of the spectral structure in more detail,
Fig.~{\ref{fig:3d_spec_6fs}(b)} shows the energy-resolved spectrum at
selected times.

The spectral structure develops gradually over several hundred
femtoseconds.
Initially, the spectrum is broad and largely unstructured, with a
maximum around \qty{0.85}{\eV}.
The first structure emerges after approximately \qty{200}{\fs} at low
kinetic energies and, over the following \qtyrange{200}{300}{\fs},
develops roughly towards higher energies.
By approximately \qty{500}{\fs}, the main spectral features are
established, and only minor changes occur thereafter.
At \qty{1}{\ps}, the spectrum has reached its final shape, discussed
above and shown in Fig.~\ref{fig:spec_t1000fs_fwhm6fs}.
This timescale is consistent with earlier theoretical simulations
\cite{Scheit03}.

\begin{figure}[!ht]
 \centering
 \includegraphics[width=\columnwidth]{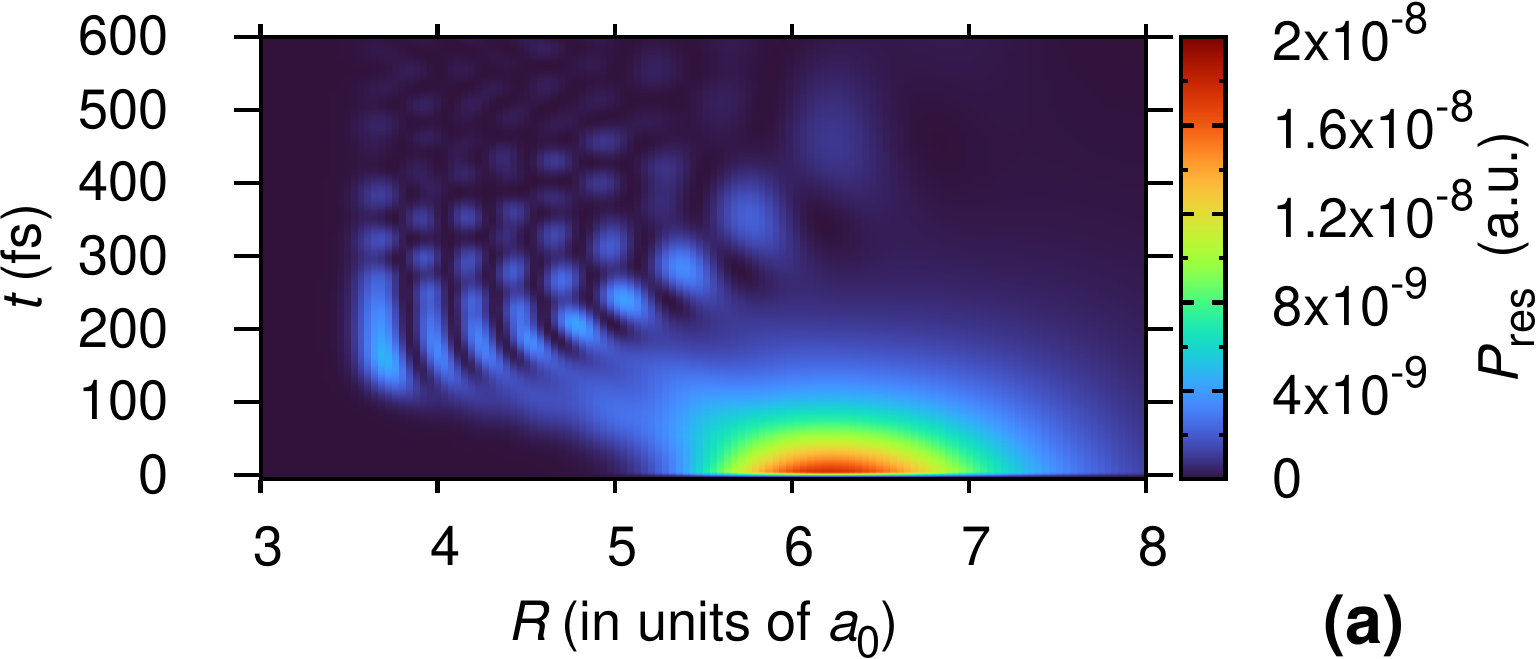}\\
 \vspace{0.5cm}
 \includegraphics[width=\columnwidth]{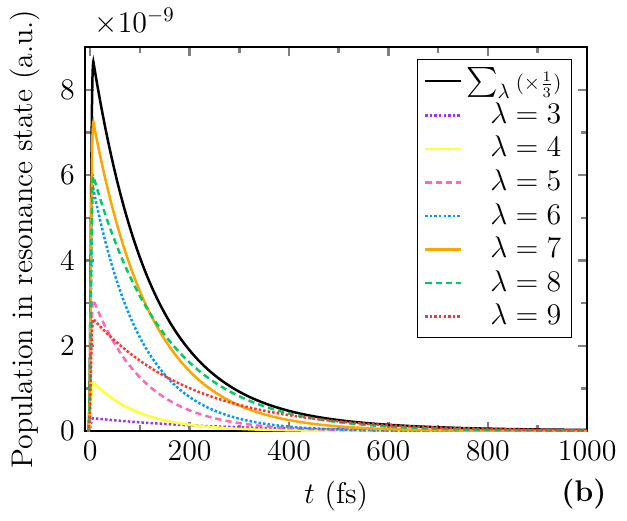}\\
 \caption{(Colour online) Evolution of the wavepacket in the electronic resonance state as a function of time $t$, at constant kinetic energy of the photoelectron, $E_p = \qty{9.754}{\eV}$. The ionizing laser pulse has FWHM of \qty{6}{\fs}.
 (a) Time- and space-resolved modulus squared. (b) Populations of the total resonance state (uppermost black curve, scaled down by a factor of 3 for better visibility) and of the dominant individual vibrational resonance states (coloured curves).
 }
 \label{fig:wavepacket_6fs}
\end{figure}

The temporal development of the ICD spectrum can be understood from
the nuclear dynamics of the resonance-state wavepacket.
Figure~\ref{fig:wavepacket_6fs}(a) shows the wavepacket as a function
of internuclear distance and time, calculated as described in
Sec.~\ref{subsec:wavepacket_res} \cite{link:suppl_lifetime_2026}.
The ionizing pulse creates the wavepacket as a coherent superposition
of vibronic resonance states, initially localized around
$R=\qty{6}{\bohr}$ in the ground state--resonance state
Franck--Condon region, see also Fig.~\ref{fig:Ne2_PEC}.
At these large internuclear distances, the wavepacket retains the
broad spatial distribution of the initial vibrational ground state
which is supported by the shallow ground-state potential.
This broad initial distribution is reflected in the broad and largely
unstructured ICD electron spectrum centred around \qty{0.85}{\eV},
consistent with the characteristic internuclear distance of about
\qty{6}{\bohr}.


The resonance-state wavepacket oscillates between smaller and larger
internuclear distances while continuously losing population through
ICD, see Fig.~{\ref{fig:wavepacket_6fs}(b)}.
Initially, it travels from the outer region of the attractive
resonance-state potential towards smaller $R$, reaching the inner
classical turning points of the bound vibronic resonance states after
approximately \qty{150}{\fs}.
This motion is also reflected in the expectation value of the
internuclear distance in the resonance state,
$\braket{R}_\text{res}
= \braket{\Psi_\text{res}|R|\Psi_\text{res}}/
\braket{\Psi_\text{res}|\Psi_\text{res}}$,
which reaches its minimum of \qty{5.23}{\bohr} at
$t=\qty{247}{\fs}$.
At the same time, the initially broad wavepacket develops pronounced
nodal structure as the coherently populated vibronic resonance states
evolve and interfere.
This coincides with the emergence of structure in the ICD electron
spectrum.


The progression of the spectral structure from low to higher kinetic
energies can be understood from an intuitive picture of the nuclear
motion.
Comparing between fixed internuclear distances, ICD becomes more efficient towards
smaller $R$.
At the same time, smaller internuclear distances correspond to lower
kinetic energies of the emitted ICD electron, as determined by the
energy difference between the resonance- and final-state potential
energy curves.
In this intuitive picture, the initial motion of the wavepacket
towards smaller $R$ therefore accounts for the early appearance of
structure at low kinetic energies.
As the wavepacket subsequently oscillates back towards larger
internuclear distances, the spectral structure continues to develop
towards higher kinetic energies.

As the decay proceeds, the continuing nuclear dynamics have a
progressively smaller effect on the time evolution of the ICD electron
spectrum.
After approximately \qty{500}{\fs}, the resonance state is already
largely depopulated and the main spectral features are well
established.
The remaining wavepacket continues to oscillate, but the resulting
variations are small compared with the signal accumulated over the
course of the decay.
By $t=\qty{1}{\ps}$, only marginal resonance-state population remains,
and the ICD electron spectrum changes only negligibly with time.
It can therefore be regarded as converged.


\begin{figure}[!ht]
 \centering
 \includegraphics[width=\columnwidth]{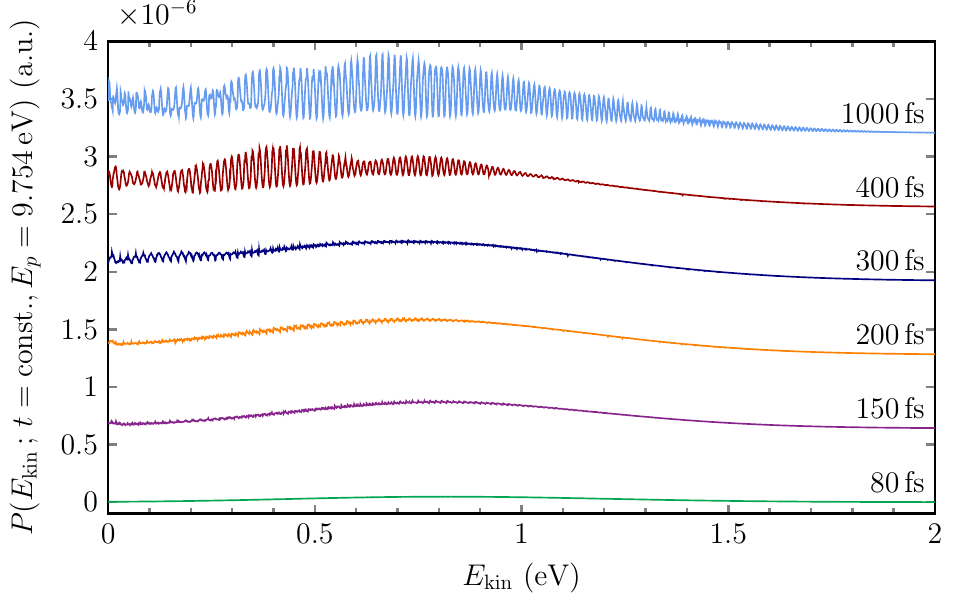}
 \caption{(Colour online) ICD signal intensity as a function of the kinetic energy of the secondary electron, $E_\text{kin}$,
 for selected constant times after the beginning of the ionizing pulse, at constant kinetic energy of the photoelectron, $E_p = \qty{9.754}{\eV}$. The ionizing laser pulse has FWHM of \qty{60}{\fs}.
 The spectra are vertically displaced against each other for better visibility.}
 \label{fig:3d_spec_60fs}
\end{figure}

\begin{figure}[!ht]
 \centering
 \includegraphics[width=\columnwidth]{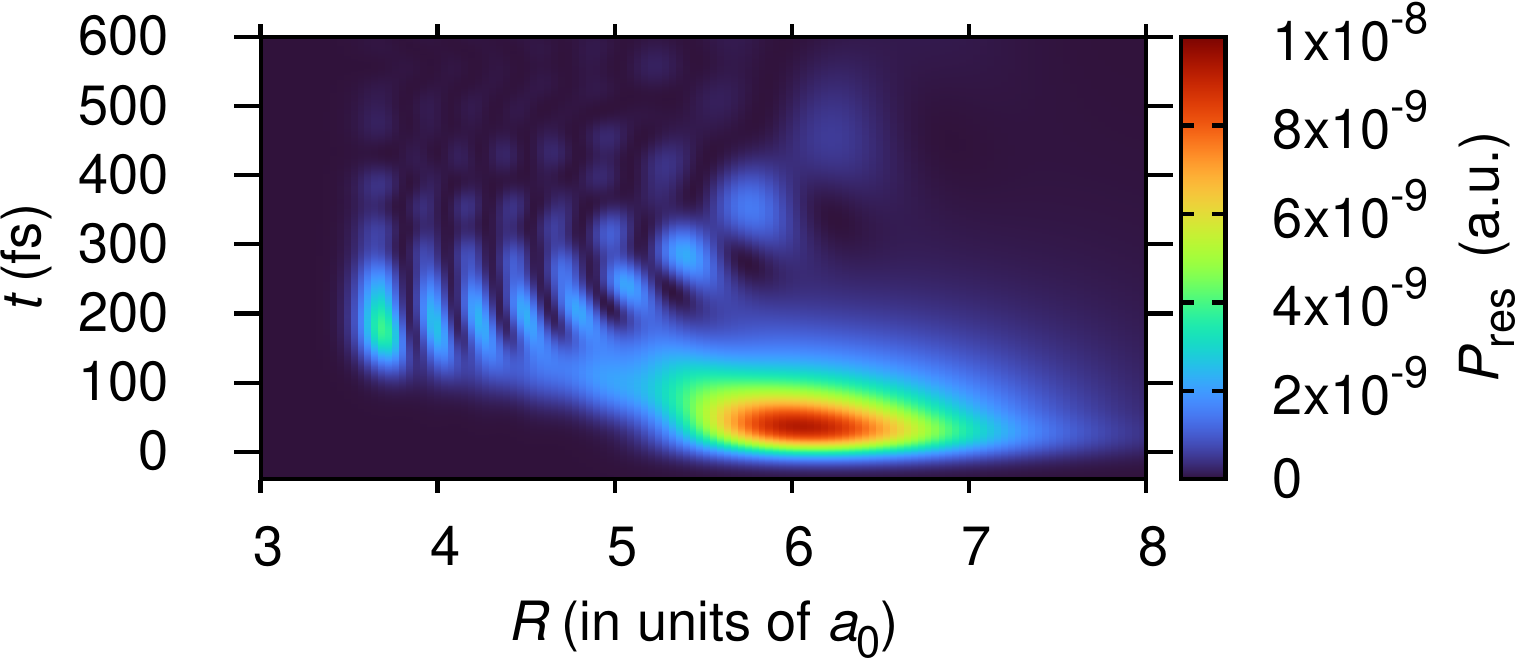}
 \caption{(Colour online) Time- and space-resolved modulus squared of the wavepacket in the electronic resonance state, at constant kinetic energy of the photoelectron, $E_p = \qty{9.754}{\eV}$. The ionizing laser pulse has FWHM of \qty{60}{\fs}.}
 \label{fig:wavepacket_60fs}
\end{figure}

The pulse duration has its strongest influence on the early-time
evolution of the ICD electron spectrum.
Figure~\ref{fig:3d_spec_60fs} shows the time- and energy-resolved
spectrum for the 60-fs pulse, while the corresponding evolution of the
resonance-state wavepacket is shown in
Fig.~\ref{fig:wavepacket_60fs} \cite{link:suppl_lifetime_2026}.
For the longer pulse, the resonance-state wavepacket builds up over
approximately \qty{50}{\fs}, during which excitation and ICD already
occur simultaneously.
This overlap is particularly relevant because the pulse duration is
comparable to the input ICD lifetime of \qty{82}{\fs}.
The early-time dynamics are therefore less clearly separated into
wavepacket preparation and subsequent decay than for the 6-fs pulse.

\subsection{Extraction of the decay lifetime}
\label{subsec:lifetime}

In time-resolved studies of electronic decay processes, decay lifetimes $\tau$
can be extracted from the temporal evolution of the decay signal.
For the neon dimer, Ref.~\cite{Schnorr15} integrated the experimental
kinetic energy release spectrum over energy and fitted the
resulting time-dependent yield with a function proportional to
$\exp(-t/\tau)$.
This procedure yielded an ICD lifetime of
\qty[separate-uncertainty = true]{150(50)}{\fs}.
Although the kinetic-energy spectrum of the ICD electron was not
measured, an equivalent analysis of this spectrum is expected to yield
the same time dependence and, consequently, the same lifetime.

Applying the same mono-exponential analysis to our simulated ICD
electron spectra, however, reveals a discrepancy with the lifetime
used as input for the simulations.
The time constant extracted from the simulated signal does not recover
the input lifetime of $\tau=\qty{82}{\fs}$, but instead lies
substantially closer to the experimentally extracted value of
\qty[separate-uncertainty = true]{150(50)}{\fs}.
The relation between the lifetime $\tau$ and the time constant
extracted from the time-resolved signal therefore requires closer
examination.



The analytical expression for the time-dependent ICD signal reveals why
the fitted time constant need not coincide with the lifetime $\tau$
that governs the decay.
Equations~\eqref{eq:ampl_working_eq}--\eqref{eq:res_ampl}, in conjunction
with Eq.~\eqref{eq:td_probability}, show that, while the wave function does obey a mono-exponential decay, the observable spectral signal for each final
state contains terms proportional to $\exp(-t/\tau)$ as well as terms
proportional to $\exp[-t/(2\tau)]$.
The temporal evolution of the observable therefore cannot, in general,
be described by a single exponential with the time constant $\tau$.
The same two temporal dependencies were previously identified in our
purely electronic description of spectator resonant ICD
\cite{Fasshauer20_1}.
A related discrepancy between a known lifetime and the temporal decay
of the signal was observed for autoionization of helium in
Ref.~\cite{Busto18}, where it was attributed to finite-pulse effects.


We therefore extract the ICD lifetime by accounting for both temporal
dependencies present in the analytical expression.
Following the experimental approach of Ref.~\cite{Schnorr15}, we first
integrate the simulated ICD electron spectrum over the kinetic energy
$E_\text{kin}$.
Starting after the pulse-induced build-up of the resonance-state
population, we fit the resulting time-dependent signal with
\begin{align}
\begin{split}
\int \mathrm{d}E_{\text{kin}}\,
P(E_{\text{kin}},t\,;\,E_p=\text{const.}) & \\
=
-a\exp\left(-\frac{t}{\tau}\right)
-b\exp\left(-\frac{t}{2\tau}\right)+c \,. &
\label{eq:int_P}
\end{split}
\end{align}
Here, the fit parameters $a$, $b$, and $c$ are positive because we follow
the population of the final states rather than the depopulation of the
decaying resonance state.
For comparison, we also fit the same signal mono-exponentially by
setting $b=0$ in Eq.~\eqref{eq:int_P}, corresponding to the procedure
used in Ref.~\cite{Schnorr15}.


\begin{figure}[!ht]
 \centering
 \includegraphics[width=\columnwidth]{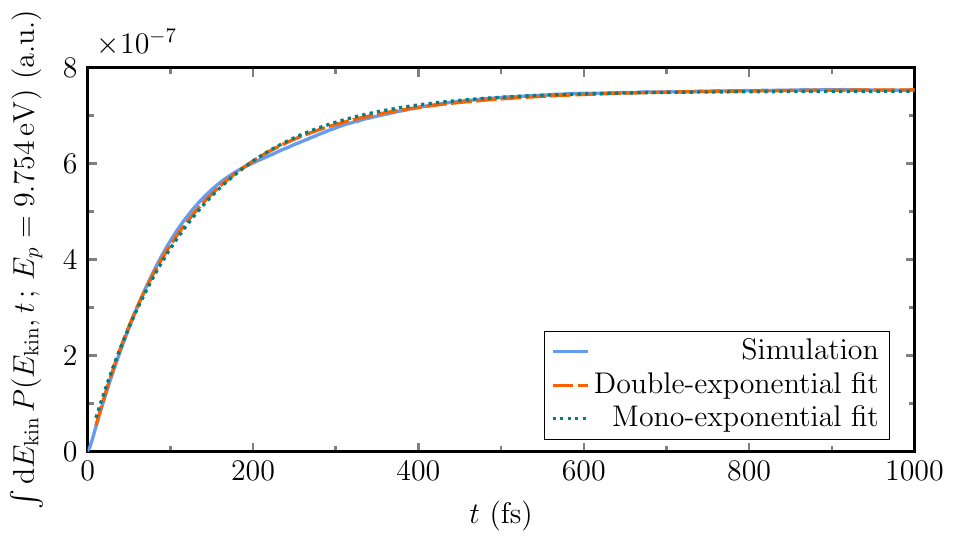}
 \caption{(Colour online) Integrated ICD signal intensity (over all values of the kinetic energy of the secondary electron, $E_\text{kin}$) (blue solid line) as a function of time, $t$, at constant kinetic energy of the photoelectron, $E_p = \qty{9.754}{\eV}$. The ionizing laser pulse has FWHM of \qty{6}{\fs}. This curve has been fitted to a double-exponential function (orange dashed line), Eq.~\eqref{eq:int_P}, and to a mono-exponential function (teal dotted line), Eq.~\eqref{eq:int_P} with $b = 0$.}
 \label{fig:int_spec_6fs}
\end{figure}

We first test the two fitting procedures against simulated data, for
which the input lifetime of $\tau=\qty{82}{\fs}$ is known.
Figure~\ref{fig:int_spec_6fs} shows the time-dependent integrated ICD
electron spectrum simulated for the 6-fs pulse together with the
mono- and double-exponential fits.
Both fits reproduce the simulated signal remarkably well, with only
minor differences visible between them.
Yet, the lifetimes they yield differ substantially.
The mono-exponential fit gives
$\tau_\text{mono}=\qty{123}{\fs}$, whereas the double-exponential fit
recovers the input lifetime with
$\tau_\text{double}=\qty{82}{\fs}$.
The quality of the fit alone therefore provides little indication of
whether the extracted time constant corresponds to the lifetime
governing the simulated decay.

%


The double-exponential fit confirms that both temporal dependencies make
substantial contributions to the simulated signal.
Their fitted weights are of similar magnitude, with
$a=\num{3.7e-7}$ and $b=\num{3.9e-7}$, while
$c=\num{7.5e-7}$.
Neither of the two exponential terms can therefore be neglected over
the fitted time interval.

\begin{figure}[t]
 \centering
 \includegraphics[width=\columnwidth]{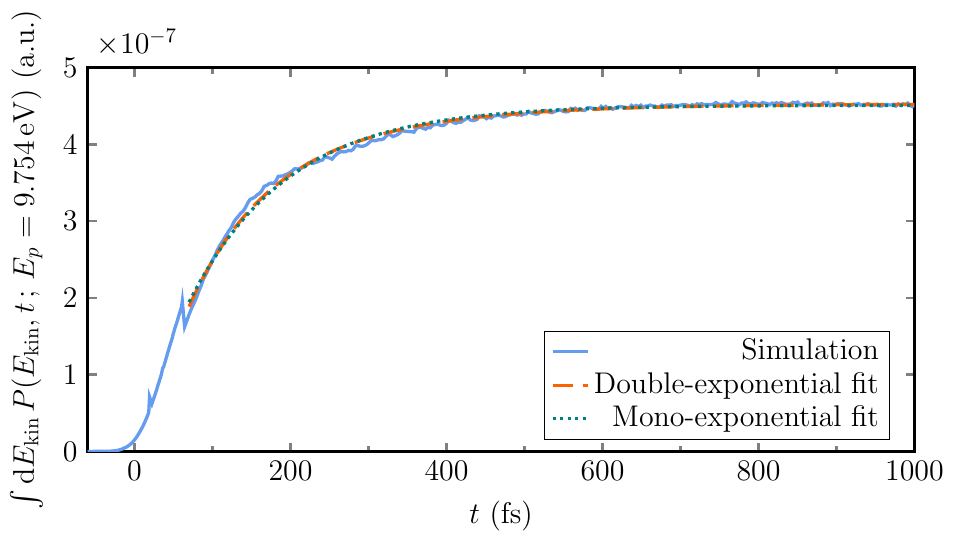}
 \caption{(Colour online) Integrated ICD signal intensity (over all values of the kinetic energy of the secondary electron, $E_\text{kin}$) (blue solid line) as a function of time, $t$, at constant kinetic energy of the photoelectron, $E_p = \qty{9.754}{\eV}$. The ionizing laser pulse has FWHM of \qty{60}{\fs}. This curve has been fitted to a double-exponential function (orange dashed line), Eq.~\eqref{eq:int_P}, and to a mono-exponential function (teal dotted line), Eq.~\eqref{eq:int_P} with $b = 0$.}
 \label{fig:int_spec_60fs}
\end{figure}

The 60-fs simulation tests the lifetime extraction under the pulse
conditions employed in the experiment.
Figure~\ref{fig:int_spec_60fs} shows the time-dependent integrated ICD
electron spectrum together with the mono- and double-exponential fits,
with the fitting interval starting at \qty{65}{\fs} to exclude the
pulse-induced build-up.
As for the 6-fs pulse, both fits reproduce the simulated signal closely
but yield substantially different lifetimes.
The mono-exponential fit gives
$\tau_\text{mono}=\qty{127}{\fs}$, close to the experimentally
extracted value of
\qty[separate-uncertainty = true]{150(50)}{\fs}
\cite{Schnorr15}, whereas the double-exponential fit yields
$\tau_\text{double}=\qty{84}{\fs}$, close to the input lifetime of
\qty{82}{\fs}.
Thus, under the experimental pulse conditions, the mono-exponential
analysis produces an apparent lifetime consistent with the previously
reported experimental value without recovering the lifetime governing
the simulated decay.

\begin{figure}[t]
 \centering
 \includegraphics[width=\columnwidth]{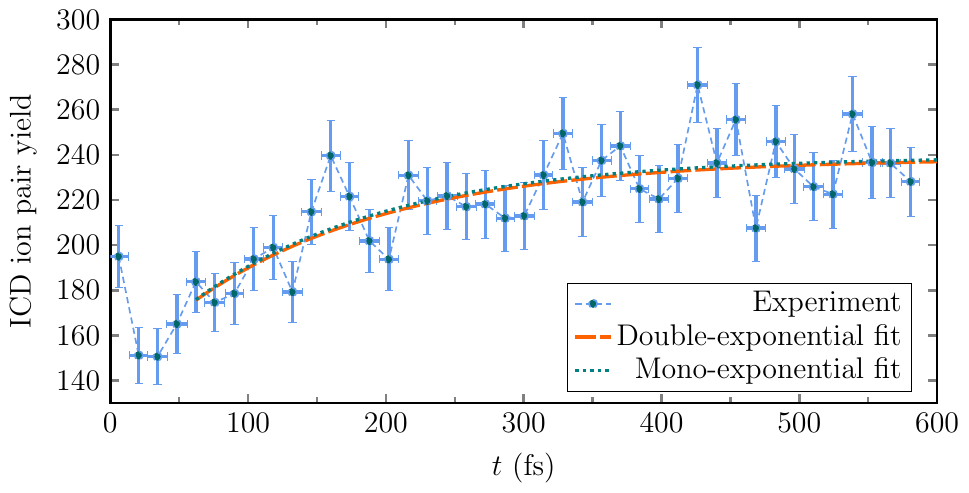}
 \caption{(Colour online) Experimentally obtained \cite{Schnorr15} integrated ICD ion pair yield (blue-green markers with uncertainty bars, fine blue dashed line meant to guide the eye) as a function of time, $t$. These data points have been fitted to a double-exponential function (orange dashed line), Eq.~\eqref{eq:int_P}, and to a mono-exponential function (teal dotted line), Eq.~\eqref{eq:int_P} with $b = 0$.}
 \label{fig:int_spec_experiment}
\end{figure}

We finally apply the same analysis to the experimental data of
Ref.~\cite{Schnorr15}, as shown in
Fig.~\ref{fig:int_spec_experiment}.
The fitting interval starts at $t=\qty{62}{\fs}$, after the end of the
pulse-induced build-up.
The mono-exponential fit yields
$\tau_\text{mono}=\qty{144}{\fs}$, reproducing the previously reported
value of \qty[separate-uncertainty = true]{150(50)}{\fs}.
By contrast, the double-exponential fit yields
$\tau_\text{double}=\qty{73}{\fs}$.
This value lies within the range of previous theoretical predictions,
which yielded lifetimes of \qty{64}{\fs} from MRCI-CAP \cite{Santra2001_4},
\qty{82}{\fs} from FanoADC-Stieltjes \cite{Averbukh06_1} and \qty{92}{\fs} from
ADC-CAP \cite{Vaval07}.
The revised extraction of the experimental lifetime thus resolves the
previous discrepancy between the experimentally extracted value and
the available theoretical predictions.

These results show that a good fit to a time-resolved decay signal is
not sufficient to identify the fitted time constant with the decay
lifetime $\tau$.
Since the underlying equations have the same form for electronic decay
processes in general, the same conclusion applies beyond ICD.
The temporal dependence of the measured observable therefore needs to
be considered when choosing the fitting function and interpreting the
resulting time constant.

\section{Conclusion}
\label{sec:conclusion}

We have extended our previous analytical description of time-resolved
electronic decay processes \cite{Fasshauer20_1,Riegel25} to include
dissociative nuclear dynamics in the final state.
The resulting framework describes the spectrum resolved in time and in
the kinetic energies of both the photoelectron and the secondary
electron, while allowing the spectrum to be evaluated directly and
independently at any given time.
This extension is essential for a realistic description of the ICD, where
the final-state nuclear dynamics generally take place in a dissociative
continuum.
We apply the framework to the neon dimer.

%

The simulated final ICD spectrum closely reproduces the spectral shape
observed experimentally in Ref.~\cite{Schnorr15} and allows us to
identify its physical origin.
A model based on the incoherent sum of contributions from individual
vibronic resonance states captures part of this shape, but neglects
interference between the different resonance pathways.
Our more coherent model reveals that these interference contributions are
essential for the spectral shape and, for the experimentally relevant
pulse duration, give rise to its dominant features.


The temporal evolution of the ICD spectrum can be understood by
relating it to the nuclear wavepacket dynamics in the decaying
electronic resonance state.
As the wavepacket oscillates while the electronic state decays, the
evolving distribution of internuclear distances is reflected in the
gradual build-up of structure in the ICD electron spectrum.


The analytical time dependence further allows us to examine directly how
the decay lifetime $\tau$ enters the time-resolved ICD spectrum.
The underlying expressions show that the integrated signal contains
terms proportional to $\exp(-t/\tau)$ as well as terms proportional to
$\exp[-t/(2\tau)]$, rather than following a single exponential with
the time constant $\tau$.
A double-exponential analysis of the simulated spectra recovers the
lifetime used as input, whereas a mono-exponential fit can reproduce
the simulated signal remarkably well without recovering this value.
The quality of the fit alone is therefore not sufficient to identify
the fitted time constant with the decay lifetime $\tau$.


Applying the same analysis to the experimental data of
Ref.~\cite{Schnorr15} yields an ICD lifetime of \qty{73}{\fs}, rather
than the previously extracted value of
\qty[separate-uncertainty = true]{150(50)}{\fs}.
The revised value lies within the range spanned by previous theoretical
predictions of \qty{64}{\fs} from MRCI-CAP, \qty{82}{\fs} from
FanoADC-Stieltjes and \qty{92}{\fs} from ADC-CAP.
Since the underlying equations have the same form for electronic decay
processes in general, the implications for lifetime extraction extend
beyond ICD.
The temporal dependence of the measured observable therefore needs to
be considered when choosing the fitting function and interpreting the
resulting time constant.

\section*{Acknowledgements}
\label{sec:acknowledgements}

E. F. and A. V. R. gratefully acknowledge funding and support through
the Core Facility LISA\textsuperscript{+} of the University of Tübingen.
The authors acknowledge support by the state of Baden-Württemberg through bwHPC
and the German Research Foundation (DFG) through grant no INST 40/575-1 FUGG (JUSTUS 2 cluster).
\par
\section*{Author Contributions}
\label{sec:author_contribs}



 \textbf{Conceptualization}: E.F.
 \textbf{Data curation}: A.V.R.
 \textbf{Formal Analysis}: A.V.R.
 \textbf{Funding}: E.F.
 \textbf{Investigation}: A.V.R., M.K.H., J.-T.K., E.F.
 \textbf{Methodology}: E.F., A.V.R.
 \textbf{Software}: A.V.R., E.F.
 \textbf{Supervision}: E.F. (lead), A.V.R. (supporting)
 \textbf{Validation}: A.V.R.
 \textbf{Visualization}: A.V.R.
 \textbf{Writing -- original draft}: A.V.R.
 \textbf{Writing -- review \& editing}: E.F., A.V.R.
\section*{Data availability}
\label{sec:DAS}
The data that support the findings of this study are openly available and cited at the appropriate locations within this paper.

\bibliographystyle{apsrev4-2}
\bibliography{theolit}
\end{document}